\documentclass[number,preprint,3p]{elsarticle}

\usepackage{nicefrac}
\usepackage{color}
\usepackage{graphicx}
\usepackage{subcaption}
\usepackage{algorithm}
\usepackage{bm}
\usepackage[colorlinks]{hyperref}% Add hyper-ref for equations, figures, etc.
\usepackage{amssymb}
\usepackage{amsthm}
\usepackage{amsmath}
\usepackage{amssymb}
\usepackage{mathtools}
\usepackage{dsfont}
\usepackage{booktabs}
\usepackage{url}
\usepackage{scrextend}
\usepackage{epstopdf}
\usepackage{float}
\usepackage{tcolorbox}
\usepackage{array}
\usepackage{tabularx}
\usepackage{booktabs}
\usepackage{multirow}
\usepackage{tablefootnote}
\usepackage[perpage]{footmisc}% Renew the footnote numbering for each page
\usepackage{bbm}
\usepackage{lineno}
\usepackage{soul}
\usepackage{caption}
\usepackage{chngcntr}

\makeatletter
\newcommand{\appendixletterlabel}[1]{%
    \begingroup
    \edef\@currentlabel{\Alph{section}}%
    \label{#1}%
    \endgroup
}
\makeatother

\DeclareUnicodeCharacter{0301}{\'{e}}

\newcommand{\vect}[1]{\boldsymbol{#1}}


\makeatletter
\gdef\urlauthor#1#2{\g@addto@macro\@elsuads{\let\corref\@gobble%
     \def\@@tmp{#1}\raggedright\eadsep
     {\ttfamily\url{\expandafter\strip@prefix\meaning\@@tmp}}\space(#2)%
     \def\eadsep{\unskip,\space}}%
}
\gdef\emailauthor#1#2{\stepcounter{ead}%
     \g@addto@macro\@elseads{\raggedright%
      \let\corref\@gobble\def\@@tmp{#1}%
      \eadsep{\ttfamily\href{mailto:\expandafter\strip@prefix\meaning\@@tmp}{\expandafter\strip@prefix\meaning\@@tmp}}
      (#2)\def\eadsep{\unskip,\space}}%
}
\makeatother
\biboptions{numbers,sort&compress,square}
\journal{arXiv}
\begin{document}
% \begin{linenumbers}
\begin{frontmatter}
\renewcommand{\thefootnote}{\fnsymbol{footnote}}
\title{Regional-scale nonlinear structural response simulation using building-specific Bouc-Wen models}
\author[1]{Sebin Oh}
\author[2]{Taeyong Kim\corref{cor1}}
\ead{tyong.kim@yonsei.ac.kr}
\cortext[cor1]{Corresponding author}
\address[1]{Department of Civil and Environmental Engineering, University of California, Berkeley, CA, United States}
\address[2]{Department of Civil and Environmental Engineering, Yonsei University, Seoul, Republic of Korea}
\begin{abstract}
  % \begin{linenumbers}
    \noindent
    Regional seismic assessments often assign structural properties by building archetype, limiting their representation of building-specific responses. This study develops a modular physics-regularized framework for identifying equivalent single-degree-of-freedom Bouc--Wen models from seismic input--response histories. A differentiable forward solver enables joint estimation of hysteretic parameters and viscous damping through an objective combining response matching with independently weighted dynamic-equilibrium and hysteretic state-evolution residuals. Before regional application, synthetic systems and four-story steel-frame hybrid simulation data are used to evaluate identification and prediction under unseen inputs, showing improved parameter estimation relative to a genetic algorithm and improved predictive performance with physics regularization, respectively. The framework is then applied to a numerical benchmark of 14,280 buildings in Milpitas, California, using models identified from simulated responses to a smaller earthquake to predict responses under larger scenarios. With the inventory, ground-motion inputs, and damage-and-loss models held fixed, archetypal and response-informed portfolios yield distinct damage-fraction distributions and substantially different structural repair costs. For the earthquake scenario with moment magnitude $M_w=6.8$, the response-informed portfolio yields a mean structural repair cost nearly twice that of the archetypal portfolio, despite nearly identical mean damage fractions. These results highlight the influence of structural model assignment on regional damage and loss estimates and the importance of incorporating building-specific response information into regional assessment.
  % \end{linenumbers}
\end{abstract}

    
\end{frontmatter}

\renewcommand{\thefootnote}{\fnsymbol{footnote}}

%% main text
\section{Introduction}
\noindent
Reliable prediction of building responses to earthquakes is essential for seismic risk assessment. Within the performance-based earthquake engineering framework, structural responses link seismic hazard to damage and loss through engineering demand parameters used to estimate repair costs, casualties, and downtime~\cite{cornell2000progress, deierlein2003framework, krawinkler2006van}. At the regional scale, response models should capture variability across the building inventory to support risk mitigation and emergency planning.

Recent studies have introduced a range of computational approaches for regional seismic response prediction and risk assessment, including deep learning-based surrogate models~\cite{kim2020pre,ning_surrogate_2026}, probabilistic simulation~\cite{oh2024long, amaya_seismic_2027,xiang_regional-scale_2022}, and methods accounting for correlations among building responses~\cite{debock2014incorporation,kang2023deep}. Despite these advances in computational efficiency, regional frameworks often assign structural properties from predefined databases, archetype models, or HAZUS-based capacity curves~\cite{lu_open-source_2020, deierlein_cloud-enabled_2020, zsarnoczay_open-source_2025, fema2024hazus}. These representations typically reflect nominal properties associated with building type, code level, and construction era. In-service buildings may depart from these assumptions because of as-built variability, deterioration, modifications or retrofits, and incomplete inventory records. The resulting discrepancies can propagate into regional damage distributions and repair-cost estimates. Incorporating building-specific response information is therefore an important complement to advances in regional simulation efficiency.

Measured seismic responses offer an opportunity to identify building-specific structural models~\cite{kerschen2006past,wang2023model}. Advances in sensing technologies, including structural monitoring using existing fiber-optic networks~\cite{SAW2025213792, liu_bridge_2026} and wireless sensor networks~\cite{rodenas_herraiz_wireless_2016}, further support the potential for scalable response data collection~\cite{ye_review_2016, lynch_summary_2006}. However, many nonlinear hysteretic identification procedures require force--displacement relationships obtained from quasi-static cyclic tests or controlled laboratory experiments, which are rarely available for ordinary in-service structures~\cite{oh_deep_2025, wang_physics-constrained_2026, li_deep_2022}. Earthquake response records instead provide ground-motion and structural-response time histories, often with incomplete response channels, measurement noise, and limited nonlinear excursion. Together with unobserved internal hysteretic states, these limitations can allow different parameter combinations to fit a calibration record yet yield divergent predictions under stronger shaking. Addressing this challenge requires using in-service sensor data to identify physics-based hysteretic models, rather than merely fitting observed responses with data-driven surrogates. Explicitly modeling restoring-force behavior and internal-state evolution supports more reliable predictions under unseen, stronger ground motions.

This study develops a modular physics-regularized framework for identifying building-specific equivalent single-degree-of-freedom (SDOF) Bouc--Wen models from seismic input--response histories. The SDOF idealization provides an efficient representation of dominant structural response for regional assessment~\cite{fema2024hazus}, while the Bouc--Wen formulation captures nonlinear hysteresis. A differentiable forward solver enables joint estimation of hysteretic parameters and viscous damping through an objective combining response matching with dynamic-equilibrium and hysteretic state-evolution residuals. The modular formulation accommodates available response channels and allows the loss terms to be weighted independently.

To assess the framework's predictive capability and its implications for regional seismic loss estimation, the proposed framework is first evaluated using synthetic Bouc--Wen systems and experimentally informed responses from a four-story steel-frame hybrid simulation. Genetic-algorithm comparisons and a complementary OpenSees-based loss ablation study assess identification performance and the contribution of physics regularization, respectively. The framework is then applied to a numerical benchmark of 14,280 buildings in Milpitas, California, evaluates models identified from simulated responses to a smaller earthquake under larger scenarios. Comparisons with archetypal Bouc--Wen models and an OpenSees multi-degree-of-freedom (MDOF) reference quantify how building-specific updating affects regional structural demands, damage distributions, and structural repair costs within the shared HAZUS-based assumptions of the portfolios.

The remainder of this paper is organized as follows. Section~\ref{sec:sdof_modeling} introduces the equivalent SDOF dynamics and Bouc--Wen model variants. Section~\ref{sec:identification} presents the identification framework and its implementation. Section~\ref{sec:method_valid} presents the synthetic and hybrid-simulation benchmarks. Section~\ref{sec:regional_benchmark} compares regional structural demands, damage, and structural repair costs. Section~\ref{sec:conclusions} summarizes the findings, limitations, and future research directions. Appendices~\ref{apdx:loss_ablation} and~\ref{apdx:portfolio_construction} provide the loss ablation study and structural portfolio construction details, respectively.

%=======================================================================
% Section 2: Response-informed identification of nonlinear SDOF models
%=======================================================================
\section{Equivalent SDOF modeling with Bouc--Wen class models}
\label{sec:sdof_modeling}
\noindent
This section presents the structural idealization underlying the proposed framework. Section~\ref{subsec:general_sdof} states the dynamics of an equivalent SDOF system with a hysteretic restoring force, and Section~\ref{subsec:bouc_wen_models} describes the Bouc--Wen class models adopted throughout this study.

%-----------------------------------------------------------------------
\subsection{Equivalent SDOF formulation for hysteretic systems}
\label{subsec:general_sdof}
\noindent
Equivalent SDOF systems are widely used in simplified procedures for estimating the seismic demands of structural components or buildings~\cite{fema440_2005, brozovic_envelope-based_2014}. Depending on the scale of the application, the generalized displacement may represent a component-level deformation, a roof displacement, or a modal coordinate of the structural system. The mass-normalized nonlinear dynamic equilibrium equation is expressed as
\begin{equation}
    \ddot{u}(t)
    + c\dot{u}(t)
    + f_s\bigl(u(t),\vect{q}(t);\vect{\theta}\bigr)
    =
    -a_g(t),
    \label{eq:nonlinear_eom}
\end{equation}
where $u$, $\dot{u}$, and $\ddot{u}$ denote the relative displacement, velocity, and acceleration, respectively; $c$ is the viscous damping coefficient per unit mass; $f_s$ is the nonlinear restoring force per unit mass; $\vect{q}$ collects the internal variables that encode the loading history; $\vect{\theta}$ represents constitutive model parameters; and $a_g$ is the ground acceleration. Throughout this study the formulation is mass-normalized, so that $c$, $f_s$, and the stiffness and strength parameters introduced below are all divided by the effective mass; $c$ accordingly has units of $\mathrm{s^{-1}}$ and $f_s$ of acceleration. This normalization reflects the fact that, without independent knowledge of the effective mass, seismic input--response data cannot distinguish the absolute mass, damping, stiffness, and strength scales.

For differential models of rate-independent hysteresis, the internal variables $\vect{q}$ evolve according to a first-order law
\begin{equation}
    \dot{\vect{q}}(t)
    =
    \vect{h}
    \bigl(
        u(t),\dot{u}(t),\vect{q}(t);\vect{\theta}
    \bigr).
    \label{eq:general_internal_state}
\end{equation}
Together, Eqs.~\eqref{eq:nonlinear_eom} and~\eqref{eq:general_internal_state} define an initial-value problem in the state $\vect{x}=[\,u\;\dot{u}\;\vect{q}^{\intercal}\,]^{\intercal}$. Models of this type include the Bouc--Wen class~\cite{bouc_forced_1967, song_generalized_2006} and the differential formulation of the Vaiana--Rosati model~\cite{vaiana_analytical_2023}; the widely used Bouc--Wen class is adopted in this study for its smooth evolution law and compact parameterization. Models whose internal state is updated by discrete rules at branch transitions, such as the piecewise-linear Ibarra--Krawinkler family~\cite{ibarra_hysteretic_2005}, do not take this form directly; extending the present formulation to such models would require an appropriate treatment of branch transitions and any additional history variables.

In the identification problem considered here, $a_g(t)$ is treated as a known input, and one or more channels among $u(t)$, $\dot{u}(t)$, and $\ddot{u}(t)$ are available as measured responses, whereas $\vect{q}(t)$ remains entirely latent. Unless otherwise specified, zero initial displacement, velocity, and internal state are assumed: $u(0)=\dot{u}(0)=0$ and $\vect{q}(0)=\vect{0}$.

%-----------------------------------------------------------------------
\subsection{Bouc--Wen class models}
\label{subsec:bouc_wen_models}
\noindent
For the Bouc--Wen models considered here, the internal-state vector $\vect{q}$ includes a dimensionless hysteretic variable $z$ that determines the hysteretic component of the restoring force. The restoring force is expressed as the sum of linear elastic
and hysteretic components,
\begin{equation}
    f_s(u,z)
    =
    \alpha k_0 u
    +
    (1-\alpha)F_y z,
    \label{eq:bw_restoring_force}
\end{equation}
where $k_0$ is the initial stiffness, $F_y$ is the nominal yield strength, $\alpha$ is the nominal ratio of post-yield to initial stiffness~\cite{kim_deep_2023}. Setting $\alpha=1$ gives a linear elastic system, whereas smaller values of $\alpha$ assign greater weight to the hysteretic component. Under the mass normalization adopted here, $k_0$ has units of $\mathrm{s^{-2}}$, and $F_y$ and $f_s$ have units of acceleration. Values of $F_y$ may also be reported as multiples of gravitational acceleration $g$. The nominal yield displacement is defined as $u_y=F_y/k_0$.

The initial stiffness and viscous damping coefficient are often expressed in terms of the initial elastic natural period $T$ and the viscous damping ratio $\zeta$ as $k_0=\omega_n^2$ and $c=2\zeta\omega_n$, respectively, where $\omega_n=2\pi/T$ is the initial elastic natural angular frequency. Substituting Eq.~\eqref{eq:bw_restoring_force} into Eq.~\eqref{eq:nonlinear_eom} gives
\begin{equation}
    \ddot{u}
    =
    -a_g
    -2\zeta\omega_n\dot{u}
    -\alpha\omega_n^2u
    -(1-\alpha)F_yz.
    \label{eq:bw_acceleration}
\end{equation}

The two variants used in this study, the basic Bouc--Wen model (BW) and the model with degradation (BWdeg), share Eq.~\eqref{eq:bw_restoring_force} but differ in the evolution of $z$. Both are described by
\begin{equation}
    \dot{z}
    =
    \frac{k_0}{F_y\eta(\varepsilon)}
    \left[
        \dot{u}
        -
        \nu(\varepsilon)
        \left(
            \beta|\dot{u}|z|z|^{n-1}
            +
            \gamma\dot{u}|z|^n
        \right)
    \right],
    \label{eq:general_bw_z_evolution}
\end{equation}
where $\beta$ and $\gamma$ control the shape of the hysteresis loop, and $n$ controls the sharpness of the transition from elastic to inelastic behavior. The functions $\eta(\varepsilon)$ and $\nu(\varepsilon)$ describe stiffness and strength degradation, respectively~\cite{oh_boucwen_2023}. The variable $\varepsilon$ is the hysteretic energy per unit mass, defined below, with units of $\mathrm{m^2/s^2}$.

We impose the normalization $\beta+\gamma=1$, so that $\gamma=1-\beta$ and only $\beta$
is estimated independently. Under monotonic loading without degradation, this normalization makes $z$ approach $1$ in the positive loading direction and $-1$ in the negative direction~\cite{oh_deep_2025}. In the positive loading direction, the initial elastic line $f_s=k_0u$ and the large-displacement asymptote $f_s=\alpha k_0u+(1-\alpha)F_y$ intersect at $(u,f_s)=(u_y,F_y)$. The parameters $F_y$ and $u_y$ therefore define the nominal yield point of the corresponding bilinear idealization.

Setting $\eta(\varepsilon)=\nu(\varepsilon)=1$ gives the basic BW model, whose internal state consists only of $z$, i.e. $\vect{q}=[z]$. The BWdeg model instead uses the linear degradation functions
\begin{equation}
    \eta(\varepsilon)
    =
    1+\delta_{\eta}\varepsilon,
    \qquad
    \nu(\varepsilon)
    =
    1+\delta_{\nu}\varepsilon,
    \label{eq:bw_degradation_functions}
\end{equation}
with the energy evolution law
\begin{equation}
    \dot{\varepsilon}
    =
    (1-\alpha)F_yz\dot{u},
    \qquad
    \varepsilon(0)=0.
    \label{eq:hysteretic_energy_evolution}
\end{equation}
Here, $\varepsilon$ represents the accumulated work per unit mass of the hysteretic component and is used to determine degradation. The internal-state vector of the BWdeg model is therefore $\vect{q}=[\,z,\varepsilon\,]^{\intercal}$.

The coefficients $\delta_{\nu}$ and $\delta_{\eta}$ govern the rates of strength and stiffness degradation with accumulated hysteretic energy, respectively, and have units of $\mathrm{s^2/m^2}$. For a fixed value of $\varepsilon$, increasing $\nu$ lowers the limiting magnitude of $z$ under monotonic loading, while increasing $\eta$ reduces the magnitude of $dz/du$ at a given $z$ and loading direction, as follows from Eq.~\eqref{eq:general_bw_z_evolution}.

In summary, the parameter vectors to be identified are
\begin{align}
    \vect{\theta}_{\mathrm{BW}}
    &=
    [\,T\;F_y\;\alpha\;\beta\;n\;\zeta\,]^{\intercal},
    \label{eq:theta_bw}
    \\
    \vect{\theta}_{\mathrm{BWdeg}}
    &=
    [\,T\;F_y\;\alpha\;\beta\;n\;
      \delta_{\nu}\;\delta_{\eta}\;\zeta\,]^{\intercal}.
    \label{eq:theta_bwdeg}
\end{align}
The BWdeg model reduces to the BW model when $\delta_{\nu}=\delta_{\eta}=0$. These two models cover the basic hysteresis and strength and stiffness degradation considered in this study. Pinching effects, which can be represented by the broader Bouc--Wen--Baber--Noori model~\cite{baber_random_1985,baber1986modeling,oh_boucwen_2023}, are not included.

%=======================================================================
% Section 3: Response-informed identification
%=======================================================================

\section{Response-informed identification}
\label{sec:identification}
\noindent
This section develops the estimation of the model parameters $\vect{\theta}$ from seismic input--response histories. Parameter estimation alternates between simulating responses and updating the parameters to minimize a physics-regularized objective. Section~\ref{subsec:differentiable_forward} describes forward simulation and gradient computation, Section~\ref{subsec:objective} defines the objective function, and Section~\ref{subsec:implementation} details the optimization procedure.

%-----------------------------------------------------------------------
\subsection{Forward simulation and gradient computation}
\label{subsec:differentiable_forward}
\noindent
Rather than directly fitting model parameters to measured restoring-force--displacement histories, the present approach identifies parameters by comparing simulated and recorded seismic responses under a known ground acceleration. Evaluating these responses and their sensitivities to the parameters requires time integration of the governing equations and differentiation through the resulting computations.

For a candidate parameter vector $\vect{\theta}$, Eqs.~\eqref{eq:nonlinear_eom} and~\eqref{eq:general_internal_state} are integrated with zero initial displacement, velocity, and internal states to obtain
\begin{equation}
    \widehat{\vect{y}}(\vect{\theta})
    =\mathcal{H}(\vect{\theta};a_g,\Delta t),
    \label{eq:forward_operator}
\end{equation}
where $\mathcal{H}$ denotes the numerical response operator and $\widehat{\vect{y}}$ collects the predicted relative displacement, velocity, and acceleration histories for the prescribed ground acceleration $a_g$ and sampling interval $\Delta t$. The operator $\mathcal{H}$ is evaluated using explicit fourth-order Runge--Kutta integration, holding $a_g$ constant within each sampling interval. As a numerical safeguard, our implementation initially uses one step per sampling interval and retries the simulation with two and then four substeps per interval if non-finite states occur. The simulation is terminated and marked as failed if non-finite states persist with four substeps. Small positive lower bounds are also imposed on $|z|$ when evaluating powers and on denominators to avoid numerical singularities.

During forward simulation, the time derivatives of the response and internal states are evaluated from the governing equations, Eqs.~\eqref{eq:nonlinear_eom} and~\eqref{eq:general_internal_state}, and integrated to obtain the response histories. Gradient-based parameter estimation additionally requires derivatives of the objective function with respect to $\vect{\theta}$, accounting for how parameter changes propagate through the evolving states and responses. To compute these gradients, the simulator is implemented in PyTorch~\cite{paszke_pytorch_2019}, which records the time-integration operations as a computational graph. Reverse-mode automatic differentiation through this graph yields the objective gradients used to update the parameters~\cite{JMLR:v18:17-468}.
%-----------------------------------------------------------------------
\subsection{Physics-regularized objective function}
\label{subsec:objective}
\noindent
The identification objective combines three loss terms: a response-matching loss $\mathcal{L}_{\mathrm{resp}}$, a dynamic-equilibrium loss $\mathcal{L}_{\mathrm{eom}}$, and a state-evolution loss $\mathcal{L}_z$. The first measures agreement with the recorded responses, while the latter two assess consistency with the governing equations.

\paragraph{Response-matching loss}
The response-matching loss normalizes discrepancies by each channel's observed root-mean-square (RMS) amplitude and averages the resulting penalties with equal weights across channels. This accounts for differences in units and response magnitudes. Let $\mathcal{C}\subseteq\{u,v,a\}$ denote the selected channels, where $u$, $v=\dot{u}$, and $a=\ddot{u}$ are the relative displacement, velocity, and acceleration, respectively. When acceleration measurements represent absolute acceleration, they are converted to relative acceleration by subtracting the ground acceleration $a_g$. For a record containing $N$ samples, the loss is defined using a pointwise penalty function $\ell$ as
\begin{equation}
    \mathcal{L}_{\mathrm{resp}}(\vect{\theta})
    =\frac{1}{|\mathcal{C}|}
    \sum_{j\in\mathcal{C}}
    \frac{1}{N}\sum_{i=1}^{N}
    \ell\!\left(
        \frac{\widehat{y}_{j}(t_i;\vect{\theta})
        -y_j^{\mathrm{obs}}(t_i)}{s_j}
    \right),
    \label{eq:response_loss}
\end{equation}
where $|\mathcal{C}|$ is the number of selected channels,
$\widehat{y}_{j}(t_i;\vect{\theta})$ and
$y_j^{\mathrm{obs}}(t_i)$ are the simulated and observed responses,
respectively, in channel $j$ at time $t_i$, and
\begin{equation}
    s_j=
    \left(
        \frac{1}{N}\sum_{i=1}^{N}
        \bigl(y_j^{\mathrm{obs}}(t_i)\bigr)^2
    \right)^{1/2}
    \label{eq:channel_scale}
\end{equation}
is the RMS amplitude of the observed response in that channel.
For the pointwise penalty $\ell$, we adopt the unit-threshold
Huber function~\cite{huber_robust_1992},
\begin{equation}
    \ell(r)=
    \begin{cases}
        \tfrac12 r^2, & |r|\leq1,\\
        |r|-\tfrac12, & |r|>1,
    \end{cases}
    \label{eq:huber}
\end{equation}
which limits the influence of large normalized residuals.

\vspace{\baselineskip}
Response matching alone may not sufficiently constrain the hysteretic parameters. When the excitation induces only limited nonlinear deformation, different parameter combinations may produce similar response histories. Two physics losses therefore assess consistency with dynamic equilibrium and the assumed state-evolution law along observed or reconstructed response histories. Evaluating these losses along the forward-simulated trajectory would largely reproduce the governing equations already enforced by time integration; the losses are evaluated using the observed responses accordingly. Below, $u$, $v$, and $a$ denote response histories obtained from measurements, with unavailable channels numerically derived from the available channels as described in Section~\ref{subsec:implementation}.

These consistency checks require the unobserved hysteretic state $z$. The dynamic-equilibrium equation, Eq.~\eqref{eq:bw_acceleration}, and the state-evolution law, Eq.~\eqref{eq:general_bw_z_evolution}, provide complementary ways to reconstruct it. Here, $\widehat{z}$ is obtained by integrating the state-evolution law along the velocity history, whereas $\widetilde{z}$ is inferred algebraically from dynamic equilibrium using the response histories defined above. These states are used to evaluate the dynamic-equilibrium and state-evolution losses, respectively. Each loss uses the state reconstructed from the other governing equation so that the residual does not vanish by construction; a state reconstructed by integrating the state-evolution law is used to evaluate the dynamic-equilibrium residual, and vice versa.

\paragraph{Dynamic-equilibrium loss}
For the dynamic-equilibrium loss $\mathcal{L}_{\mathrm{eom}}$, the internal evolution law is integrated along the prescribed velocity history $v$, $\dot{\widehat{z}}=h(v,\widehat{z},\widehat{\varepsilon};\vect{\theta})$, starting from $\widehat{z}(0)=0$, where $h$ denotes the right-hand side of Eq.~\eqref{eq:general_bw_z_evolution}. For the BWdeg model, the associated hysteretic energy is integrated simultaneously as $\dot{\widehat{\varepsilon}}=(1-\alpha)F_y\widehat{z}v$, with $\widehat{\varepsilon}(0)=0$, according to Eq.~\eqref{eq:hysteretic_energy_evolution}. The resulting equilibrium residual is
\begin{equation}
    r_{\mathrm{eom}}
    =a+a_g+2\zeta\omega_n v
    +\alpha\omega_n^2u+(1-\alpha)F_y\widehat{z},
    \label{eq:eom_residual}
\end{equation}
which measures how closely dynamic equilibrium is satisfied when the hysteretic state is reconstructed from the prescribed response histories. Note that replacing $\widehat{z}$ with the equilibrium-based state $\widetilde{z}$ defined below would make this residual identically zero by construction.

To reduce sensitivity to high-frequency measurement noise, we average the residual over overlapping windows of $w$ samples. The $i$th output component of the moving-average operator $\mathcal{A}_w$ is defined as
\begin{equation}
    \mathcal{A}_w[r_{\mathrm{eom}}]_i
    =\frac{1}{w}\sum_{k=i}^{i+w-1}r_{\mathrm{eom}}(t_k),
    \qquad i=1,\ldots,N-w+1.
    \label{eq:weak_form}
\end{equation}
The dynamic-equilibrium loss is then
\begin{equation}
    \mathcal{L}_{\mathrm{eom}}(\vect{\theta})
    =\frac{1}{N-w+1}\sum_{i=1}^{N-w+1}
    \ell\!\left(
        \frac{\mathcal{A}_w[r_{\mathrm{eom}}]_i}{F_y}
    \right).
    \label{eq:eom_loss}
\end{equation}
Dividing the residual by $F_y$ makes it dimensionless and measures the force imbalance as a fraction of the nominal yield strength. The denominator is updated to the current value of $F_y$ at each iteration but is treated as constant during gradient computation to avoid biasing the optimization toward larger values of $F_y$.

\paragraph{State-evolution loss}
For the state-evolution loss $\mathcal{L}_{\mathrm{z}}$, we use an integral form of the state-evolution equation to avoid amplifying measurement noise through numerical differentiation of the reconstructed hysteretic state. The internal state is first inferred from dynamic equilibrium as

\begin{equation}
    \widetilde{z}
    =-\frac{a+a_g+2\zeta\omega_n v+\alpha\omega_n^2u}
    {(1-\alpha)F_y}.
    \label{eq:z_reconstruction}
\end{equation}
For BWdeg, the associated energy history is calculated from
\begin{equation}
    \widetilde{\varepsilon}(t)
    =\int_0^t (1-\alpha)F_y\widetilde{z}(\tau)v(\tau)
    \,\mathrm{d}\tau,
    \label{eq:energy_reconstruction}
\end{equation}
according to Eq.~\eqref{eq:hysteretic_energy_evolution}.
The integral is evaluated using the trapezoidal rule with $\widetilde{\varepsilon}(0)=0$.

To avoid numerical differentiation of the reconstructed hysteretic state $\widetilde{z}$, which can amplify measurement noise and reconstruction errors, the state-evolution equation is integrated over overlapping windows of $w$ time intervals rather than evaluating the differential residual $\dot{\widetilde{z}}-h(v,\widetilde{z},\widetilde{\varepsilon};\vect{\theta})$. This formulation compares the reconstructed state increment with the accumulated evolution predicted by the model, yielding
\begin{equation}
    r_{\mathrm{z},i}
    =\widetilde{z}_{i+w}-\widetilde{z}_i
    -\Delta t\sum_{k=i}^{i+w-1}
    h(v_k,\widetilde{z}_k,\widetilde{\varepsilon}_k;\vect{\theta})
    \label{eq:evolution_residual}
\end{equation}
and
\begin{equation}
    \mathcal{L}_{\mathrm{z}}
    =\frac{1}{N-w}\sum_{i=1}^{N-w}
    \ell\!\left(\frac{r_{\mathrm{z},i}}{s_z}\right),
    \qquad
    s_z=\max_{1\leq i\leq N}|\widetilde{z}_i|.
    \label{eq:z_loss}
\end{equation}
The residual is normalized by $s_z$ to express the state-evolution discrepancy relative to the peak amplitude of the reconstructed hysteretic state. The denominator $s_z$ is updated at each iteration, and its dependence on $\vect{\theta}$ is retained during gradient computation to account for changes in the residual relative to the amplitude of the reconstructed hysteretic state.

\paragraph{Loss balancing}
Each active loss term is normalized by its value at the initial parameter guess $\vect{\theta}_0$. The composite objective is
\begin{equation}
    \mathcal{J}(\vect{\theta})=\lambda_{\mathrm{resp}}\frac{\mathcal{L}_{\mathrm{resp}}(\vect{\theta})}{\mathcal{L}_{\mathrm{resp}}(\vect{\theta}_0)}+\lambda_{\mathrm{eom}}\frac{\mathcal{L}_{\mathrm{eom}}(\vect{\theta})}{\mathcal{L}_{\mathrm{eom}}(\vect{\theta}_0)}+\lambda_z\frac{\mathcal{L}_z(\vect{\theta})}{\mathcal{L}_z(\vect{\theta}_0)}.
    \label{eq:composite_objective}
\end{equation}
These scales remain fixed throughout optimization. Unless otherwise stated, equal weights are used through out this study: $\lambda_{\mathrm{resp}}=\lambda_{\mathrm{eom}}=\lambda_z=1$. Ablation studies set the omitted terms' weights to zero. The combined objective balances response matching with consistency with the governing equations but does not guarantee unique parameter recovery. Its effects on variability across calibration records and predictive performance on unseen records are assessed in Section~\ref{sec:method_valid}.

%-----------------------------------------------------------------------
\subsection{Implementation details}
\label{subsec:implementation}
\paragraph{Channels and record preparation}
Response channels required for the physics losses are reconstructed when unavailable. Velocity and acceleration are obtained from displacement by successive numerical differentiation, using central differences at interior samples. When only acceleration is available, velocity and displacement are obtained by successive numerical integration using the forward rectangle rule, assuming zero initial velocity and displacement.

The computational cost of parameter estimation increases with the duration of the response record, motivating the removal of unnecessary data points. Moreover, long low-amplitude tails can also disproportionately influence the time-averaged losses, potentially favoring agreement with these portions over the strong-motion response. For earthquake records, we therefore retain the record from its original start through two seconds after the cumulative Arias intensity of the ground acceleration reaches $99\%$ of its total value, or to the end of the record if it occurs sooner.

\paragraph{Bounds and optimization}
The parameter bounds are adopted from the literature~\cite{oh_deep_2025} and summarized in Table~\ref{tab:bounds}. To account for differences in parameter scales, each parameter $\theta_j$ is expressed in terms of a dimensionless, unconstrained optimization variable $\phi_j$ through the sigmoid transformation
\begin{equation}
\theta_j=\theta_j^{\min}+(\theta_j^{\max}-\theta_j^{\min})\frac{1}{1+\exp(-\phi_j)},
\label{eq:bounded_parameter_map}
\end{equation}
which keeps the parameter within its prescribed bounds. We adopt Adam~\cite{kingma2017adammethodstochasticoptimization} to update the optimization variables $\vect{\phi}$. The gradient norm is clipped at $10$. A learning rate of $0.001$ or $0.002$ is used and held constant throughout each fit. The final iterate is returned as the parameter estimate $\vect{\theta}^{*}$. Convergence can be assessed using indicators such as the relative change in the mean total loss between successive windows of training epochs and the stability of the parameter estimates. The window length and convergence tolerances can be selected based on the application requirements and the available computational budget.

\begin{table}[H]
\centering
\caption{\textbf{Parameter bounds for mass-normalized identification.}}
\label{tab:bounds}
\begin{tabular}{llccl}
\hline
Parameter & Symbol & Lower & Upper & Unit \\
\hline
Natural period                    & $T$            & 0.05 & 3.00 & s \\
Nominal yield strength            & $F_y$          & 0.05 & 1.00 & $g$ \\
Post-to-pre-yield stiffness ratio  & $\alpha$       & 0.01 & 0.50 & -- \\
Hysteresis loop shape             & $\beta$        & 0.10 & 0.90 & -- \\
Transition sharpness              & $n$            & 1.00 & 5.00 & -- \\
Strength-degradation rate         & $\delta_{\nu}$ & 0.00 & 0.50 & $\mathrm{s^{2}/m^{2}}$ \\
Stiffness-degradation rate        & $\delta_{\eta}$& 0.00 & 0.50 & $\mathrm{s^{2}/m^{2}}$ \\
Viscous damping ratio             & $\zeta$        & 0.00 & 0.15 & -- \\
\hline
\end{tabular}
\end{table}

%=======================================================================
% Section 4: Methodological validation
%=======================================================================
\section{Methodological validation}
\label{sec:method_valid}
\noindent
Before examining regional loss estimates, this section evaluates parameter recovery and response prediction in two benchmarks. The synthetic BWdeg benchmark (Section~\ref{subsec:valid_synthetic}) compares the proposed gradient-based algorithm with a genetic algorithm (GA) under comparable elapsed fitting times, without model-form mismatch. The hybrid-simulation application (Section~\ref{subsec:valid_hybrid}) assesses identification from experimentally informed structural responses. Both benchmarks evaluate predictions under unseen excitations. A complementary loss ablation study in Appendix~\ref{apdx:loss_ablation} examines the contribution of physics regularization under model-form mismatch, showing reductions in mean displacement prediction error and variability across calibration records.

%-----------------------------------------------------------------------
\subsection{Parameter recovery in synthetic BWdeg systems}
\label{subsec:valid_synthetic}
\noindent
This benchmark evaluates the ability of the proposed algorithm to recover prescribed BWdeg parameters and predict nonlinear responses under unseen inputs. The reference and calibrated models share the BWdeg formulation described in Section~\ref{subsec:bouc_wen_models}, allowing estimation performance to be evaluated without model-form mismatch. A GA provides a derivative-free baseline using the same displacement, velocity, and relative acceleration observations.

The reference systems are defined by 100 parameter sets generated using eight-dimensional Latin hypercube sampling. Each system has unit mass, and both methods use the same admissible estimation bounds in Table~\ref{tab:bounds}. Gradient-based optimization uses common initial values across all systems for the hysteretic and damping parameters: $\alpha=0.105$, $\beta=0.50$, $n=2.00$, $\delta_\nu=\delta_\eta=0.20~\mathrm{s^2/m^2}$, and $\zeta=0.05$. A different initialization is adopted for $T$ and $F_y$ to emulate applications in which approximate prior information on period and yield strength is available~\cite{fema2024hazus}. Their initial guesses are randomly generated for each system to differ from the prescribed values by 15--35\% and 20--40\%, respectively. Minimum deviations of 15\% for $T$ and 20\% for $F_y$ are imposed to avoid overly favorable initial guesses for the optimizer. The GA initial population follows the same perturbation rule for $T$ and $F_y$, while its remaining parameters are sampled uniformly over the full admissible estimation bounds.

\begin{figure}[H]
    \centering
    \includegraphics[width=0.60\linewidth]{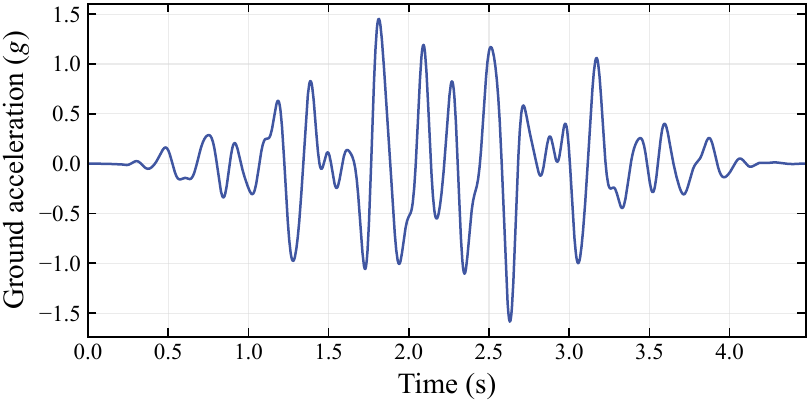}
    \caption{\textbf{Example multisine ground acceleration used for parameter estimation.} The waveform corresponds to one reference system; frequencies, duration, and amplitude are scaled according to the system period and the yield strength.}
    \label{fig:estimation_input_theta_0001}
\end{figure}

For each system, both methods use the same response histories generated under a single multisine ground-acceleration input (Figure~\ref{fig:estimation_input_theta_0001}). To induce pronounced hysteretic responses for parameter identification, input frequencies and duration are scaled according to the system period, while the amplitude is scaled according to its yield strength. Both methods use the same three-term objective comprising the response-matching ($\mathcal{L}_{\mathrm{resp}}$), dynamic-equilibrium ($\mathcal{L}_{\mathrm{eom}}$), and state-evolution ($\mathcal{L}_z$) losses. The proposed gradient-based method uses 6,000 epochs. GA uses a population of 100 and an elapsed fitting-time budget of 14 hours per system, matching the median elapsed fitting time of the gradient-based runs. Computations are performed on the \texttt{savio2\_htc} partition of Savio, the high-performance computing cluster at the University of California, Berkeley, using 100 parallel workers, with one CPU core assigned to each system.

\begin{figure}[H]
    \centering
    \includegraphics[width=\linewidth, trim=0cm 0.3cm 0cm 0cm, clip]{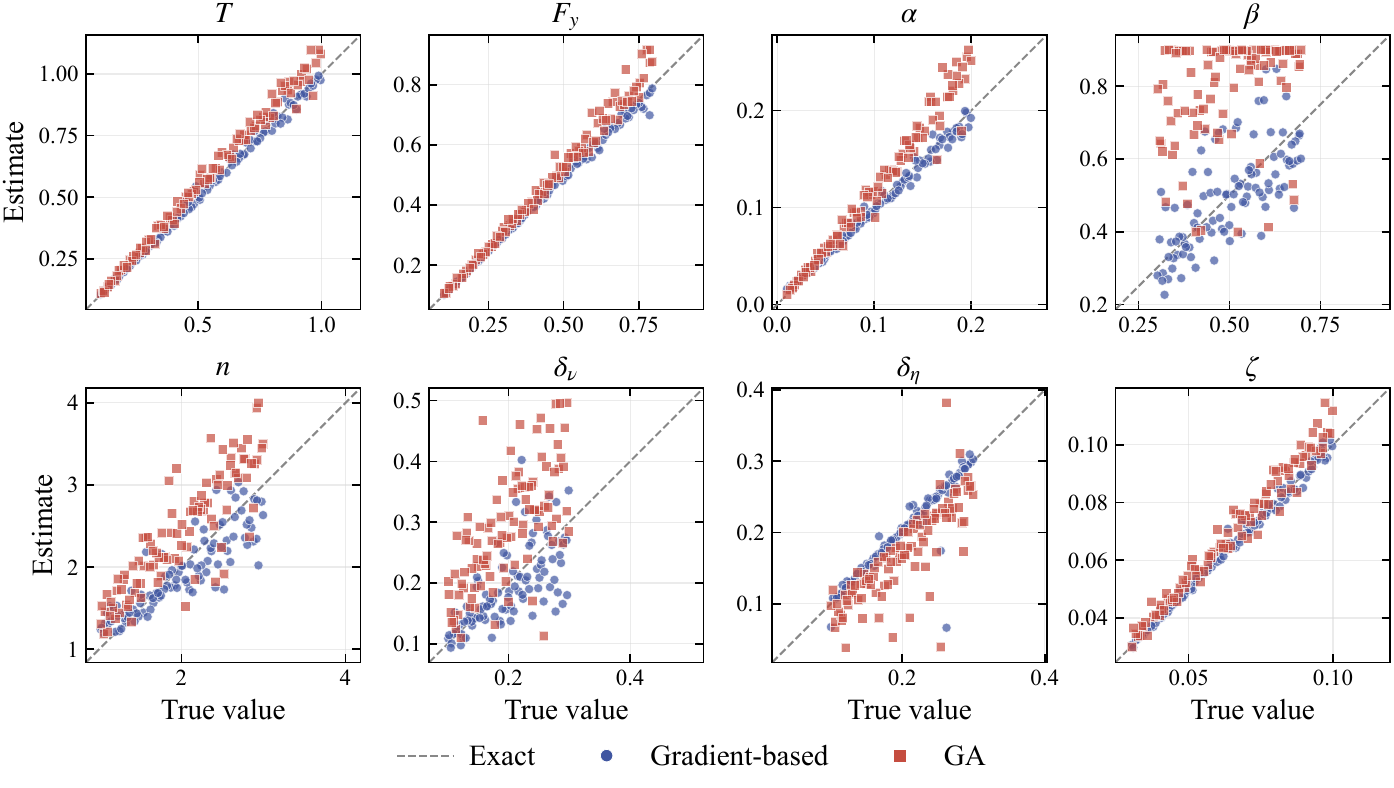}
    \caption{\textbf{Recovery of prescribed BWdeg parameters by the proposed gradient-based algorithm and GA.} Each panel compares estimates with true values for 100 synthetic BWdeg systems, with both methods using the same displacement, velocity, and relative acceleration observations. Blue circles denote the proposed method after 6,000 epochs; red squares denote the best GA candidates obtained within a 14-hour fitting budget per system, matching the median elapsed fitting time of the gradient-based runs. Dashed lines indicate exact recovery. $T$ is expressed in seconds, yield strength is shown as $F_y/g$, and the degradation coefficients are expressed in $\mathrm{s^2/m^2}$; the remaining parameters are dimensionless.}
    \label{fig:parameter_recovery_proposed_vs_ga}
\end{figure}

Figure~\ref{fig:parameter_recovery_proposed_vs_ga} shows that the proposed algorithm achieves lower median absolute relative errors for all eight parameters. Summary statistics for the absolute relative errors are provided in Table~\ref{tab:parameter_recovery_proposed_vs_ga}. The estimates of $T$, $F_y$, $\alpha$, $\delta_\eta$, and $\zeta$ closely follow the exact-recovery lines, with median errors of approximately 1.3--3.5\%. Here, absolute relative error is the absolute difference between an estimate and its true value, divided by that true value. GA captures the overall variation in period, strength, and damping, but its estimates of $T$, $F_y$, and $\alpha$ exhibit systematic overestimation biases, particularly pronounced for $\alpha$. The median signed relative errors in $T$ and $F_y$ are approximately $+8.4\%$ and $+4.8\%$, respectively. Its recovery of the other hysteretic and degradation parameters is also limited. For example, the median absolute relative errors in $\beta$ and $\delta_\nu$ are $60.4\%$ and $48.1\%$ for GA, compared with $12.0\%$ and $12.1\%$ for the proposed method. For $\beta$, $n$, and $\delta_\nu$, gradient-based estimation substantially improves parameter recovery under comparable elapsed fitting times. The damping ratio is also identified jointly with the hysteretic parameters, with a median absolute relative error of $1.3\%$ for the proposed method, compared with $6.8\%$ for GA.

\begin{table}[H]
    \centering
    \caption{\textbf{Statistics of the parameter-recovery errors for the 100 BWdeg systems.} For each parameter and case, the error is defined as the absolute relative error. The reported values are the mean, median, and sample standard deviation (Std. dev.) of these errors across the 100 cases, expressed as percentages.}
    \label{tab:parameter_recovery_proposed_vs_ga}
    \small
    \begin{tabular}{crrrrrr}
        \toprule
        & \multicolumn{3}{c}{Gradient-based}
        & \multicolumn{3}{c}{GA} \\
        \cmidrule(lr){2-4}
        \cmidrule(lr){5-7}
        Parameter
        & Mean error
        & Median error
        & Std. dev.
        & Mean error
        & Median error
        & Std. dev. \\
        \midrule
        $T$
        &  2.19 &  1.54 &  1.88
        &  8.40 &  8.38 &  3.84 \\
        $F_y$
        &  1.86 &  1.54 &  1.57
        &  5.90 &  4.77 &  4.82 \\
        $\alpha$
        &  6.11 &  3.48 &  8.38
        & 18.15 & 17.61 &  8.54 \\
        $\beta$
        & 16.00 & 12.00 & 13.58
        & 69.12 & 60.36 & 39.81 \\
        $n$
        & 10.85 &  8.88 &  8.89
        & 25.11 & 23.57 & 13.60 \\
        $\delta_{\nu}$
        & 16.90 & 12.12 & 14.34
        & 51.10 & 48.11 & 33.93 \\
        $\delta_{\eta}$
        &  3.62 &  1.37 &  8.81
        & 18.10 & 13.70 & 16.02 \\
        $\zeta$
        &  1.89 &  1.29 &  1.78
        &  7.36 &  6.82 &  4.65 \\
        \bottomrule
    \end{tabular}
\end{table}

\begin{figure}[H]
    \centering
    \includegraphics[width=0.8\linewidth]{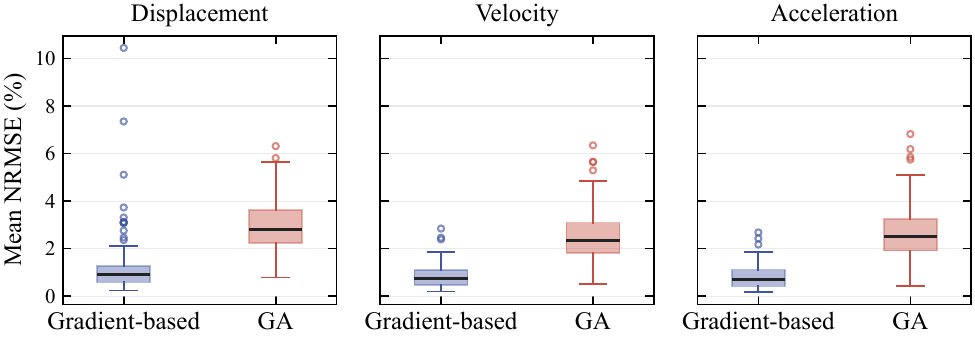}
    \caption{\textbf{Response prediction errors under unseen multisine inputs.} Each boxplot summarizes NRMSE values for 100 synthetic systems, with each value averaged over ten unseen inputs for one system. Blue boxes denote the proposed gradient-based method and red boxes denote GA. Boxes span the interquartile range (IQR), center lines denote medians, whiskers extend to the most extreme values within 1.5 IQR of the box boundaries, and circles denote outliers.}
    \label{fig:validation_nrmse}
\end{figure}

Predictive performance is assessed under the ten unseen inputs, applying the same input to the reference and both calibrated models for each system. For each response channel, the normalized root-mean-square error (NRMSE) is calculated by dividing the root-mean-square prediction error by the root-mean-square reference response, and then averaged over the ten inputs for each system. Figure~\ref{fig:validation_nrmse} compares these averages across the 100 systems. The median NRMSE values for the proposed method are $0.91\%$, $0.74\%$, and $0.69\%$ for displacement, velocity, and relative acceleration, respectively. A few displacement outliers exhibit larger errors for the proposed method than for GA, with time-history inspection suggesting that discrepancies in late-time residual displacement contribute to these errors. Further tuning of the training duration and loss weights may help reduce these discrepancies and improve residual displacement prediction. Overall, the proposed method achieves median NRMSE values below $1\%$ in all three channels, approximately one-third of the corresponding GA values, while providing more accurate parameter estimates under comparable elapsed fitting times.

%-----------------------------------------------------------------------
\subsection{Hybrid simulation}
\label{subsec:valid_hybrid}
\noindent
Mortazavi et al.~\cite{mortazavi_pseudodynamic_2023} conducted pseudo-dynamic hybrid simulations of a four-story steel eccentrically braced frame (EBF) representing an office building in downtown Los Angeles and equipped with cast steel replaceable modular yielding links. The first-story CML100 link was physically tested under combined shear and flexure with negligible axial loading, while the remaining structure was modeled in OpenSees. The response histories used here were obtained under scaled ground motions from the L'Aquila, Northridge, and Kobe earthquakes and are publicly available through Harvard Dataverse~\cite{DVN/HR5DR2_2025}.

A mass-normalized BWdeg model is fitted to the Kobe roof response for 3,000 training epochs using both displacement and acceleration observations. The dataset provides story drift ratios and floor accelerations but no independent velocity histories. Roof displacement relative to the ground is reconstructed as $u(t)=\sum_{i=1}^{4}h_i\,\mathrm{SDR}_i(t)/100$, where $h_i=3.7~\mathrm{m}$ is the height of story $i$ and $\mathrm{SDR}_i(t)$ is its interstory drift ratio in percent. Roof acceleration is converted to relative acceleration by subtracting the ground acceleration. An effective participation factor is calculated as $\Gamma=\sum_{i=1}^{4}m_i\phi_i/\sum_{i=1}^{4}m_i\phi_i^2$, using the reported floor masses $m_i$ and effective first-mode shapes estimated from the measured floor displacement histories. Each mode shape is normalized by its roof component such that $\phi_4=1$. The resulting value, $\Gamma \approx 1.36$, is used throughout the calibration and prediction analyses.

Two loss configurations are compared using the same initial parameters and optimization settings. The common initial values are $T=0.843~\mathrm{s}$, $F_y=0.305g$, $\alpha=0.142$, $\beta=0.315$, $n=2.076$, $\delta_\nu=\delta_\eta=0.134~\mathrm{s^2/m^2}$, and $\zeta=0.0403$. Response-only fitting uses $(\lambda_{\mathrm{resp}},\lambda_{\mathrm{eom}},\lambda_z)=(1,0,0)$, whereas physics-regularized fitting uses $(1,0.5,0.2)$ in Eq.~\eqref{eq:composite_objective}. The reduced weights on the physics terms reflect the expected model-form mismatch when representing a multistory frame and its distributed nonlinear behavior using a single BWdeg oscillator. These illustrative weights are selected by judgment, without a search for optimal values, demonstrating the modular weighting of the loss terms.

\begin{figure}[H]
    \centering
    \includegraphics[width=0.8\linewidth, trim=0cm 0.3cm 0cm 0.4cm, clip]{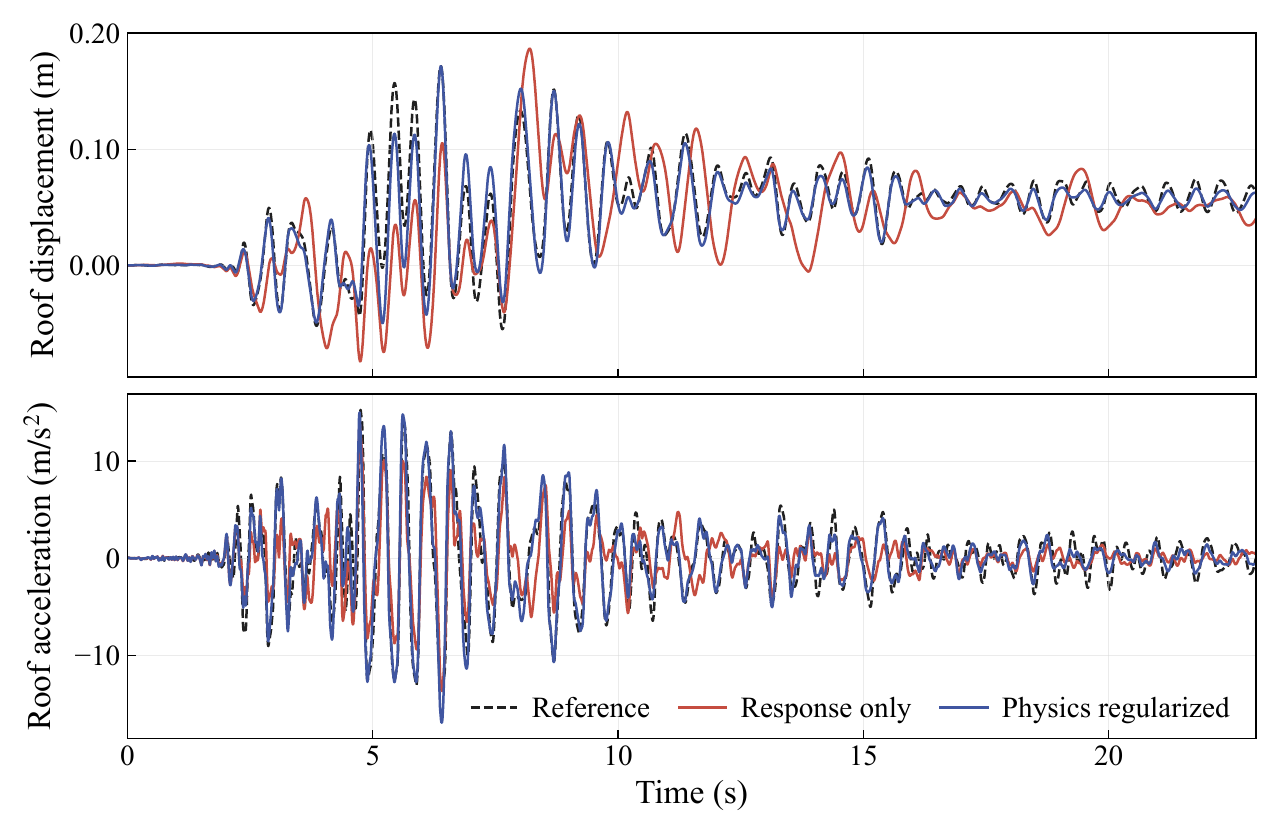}
    \caption{\textbf{Response reproduction for the Kobe hybrid simulation record.} Roof displacement (top) and relative acceleration (bottom) reproduced using BWdeg parameters identified using roof displacement and relative acceleration observations over 3,000 epochs. Response-only and physics-regularized fitting use loss weights $(1,0,0)$ and $(1,0.5,0.2)$, respectively. Dashed curves denote the hybrid simulation reference responses.}
    \label{fig:hybrid_calibration}
\end{figure}

Figure~\ref{fig:hybrid_calibration} compares the responses reproduced using the two calibrated parameter sets. Physics regularization improves agreement with the Kobe reference histories in both channels, including the displacement offset during the later part of the record. The displacement NRMSE decreases from $61.9\%$ for response-only fitting to $18.0\%$ with physics regularization, while the acceleration NRMSE decreases from $66.5\%$ to $34.5\%$. Some acceleration peaks and short-period fluctuations remain imperfectly reproduced, consistent with the limitations of the reduced-order representation.

\begin{figure}[H]
    \centering
    \includegraphics[width=0.8\linewidth, trim=0cm 0.2cm 0cm 0.2cm, clip]{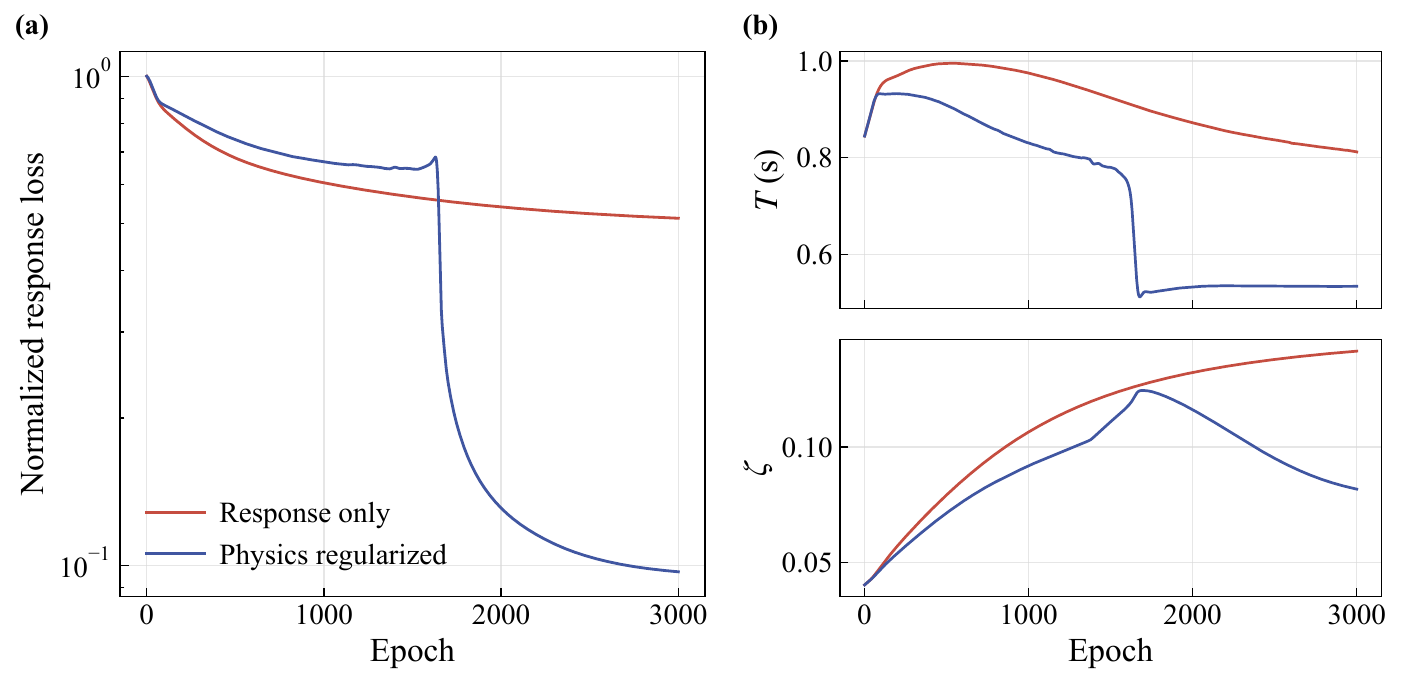}
    \caption{\textbf{Training histories for calibration to the Kobe record.} (a) Response loss normalized by its value at the common initial parameter guess, plotted on a logarithmic vertical scale. (b) Estimates of $T$ (top) and $\zeta$ (bottom) over 3,000 epochs. Both configurations use the same roof displacement and relative acceleration observations; the curves in (a) show only the response-loss component, excluding the physics terms.}
    \label{fig:hybrid_training_histories}
\end{figure}

The training histories in Figure~\ref{fig:hybrid_training_histories} clarify how the two fits diverge. Although response-only fitting initially reduces the response loss more rapidly, the physics-regularized trajectory undergoes a pronounced reduction around 1,650 epochs and reaches a substantially lower response loss. This transition coincides with a sharp decrease in $T$ and a subsequent decrease in $\zeta$. At 3,000 epochs, the relative changes in mean total loss between the final two 50-epoch windows are $0.18\%$ and $0.30\%$ for response-only and physics-regularized fitting, respectively. The final physics-regularized estimates are $T=0.534~\mathrm{s}$ and $\zeta=0.082$, compared with $T=0.812~\mathrm{s}$ and $\zeta=0.142$ for response-only fitting. The former period is consistent with a rough estimate of $0.5$--$0.6~\mathrm{s}$ obtained from the slopes of approximately linear portions of the observed displacement--acceleration curves. This estimate provides a qualitative reference for the identified period. The histories suggest that the added physics terms guide optimization toward parameters with better response agreement, without establishing unique parameter recovery or a globally optimal solution.

\begin{figure}[H]
    \centering
    \includegraphics[width=0.8\linewidth, trim=0cm 0.3cm 0cm 0.4cm, clip]{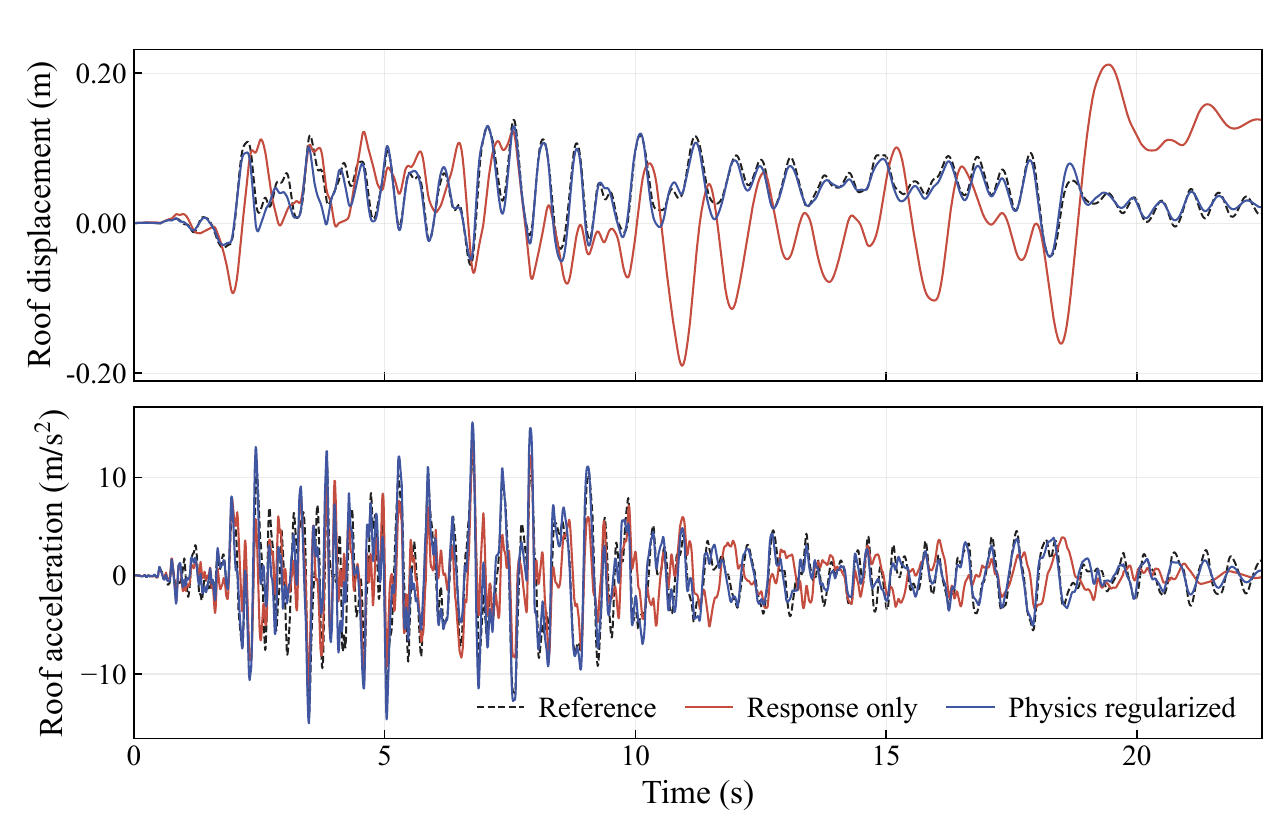}
    \caption{\textbf{Response prediction for the unseen Northridge record.} Roof displacement (top) and relative acceleration (bottom) predicted using the response-only and physics-regularized BWdeg parameter sets identified from the Kobe roof displacement and relative acceleration responses after 3,000 epochs, without further calibration. Dashed curves denote the hybrid simulation reference responses.}
    \label{fig:hybrid_prediction}
\end{figure}

The two parameter sets identified from Kobe are subsequently applied to the unseen Northridge excitation without further calibration (Figure~\ref{fig:hybrid_prediction}). The physics-regularized model more closely captures the displacement oscillations and the residual displacement near the end of the record, whereas the response-only model develops large phase and displacement-offset errors. Displacement NRMSE decreases from $160.0\%$ to $19.4\%$, and acceleration NRMSE decreases from $86.8\%$ to $52.3\%$. The improvement across both displacement and acceleration channels shows that the benefit of physics regularization in this comparison extends beyond the calibration record.

This example demonstrates the applicability of the proposed identification framework to experimental hybrid simulation data for a four-story steel frame, extending its validation beyond numerical examples. Despite the model-form mismatch associated with the reduced-order representation, physics regularization improves both calibration accuracy and prediction under an unseen earthquake record.

%=======================================================================
% Section 5: Regional-scale building portfolio benchmark
%=======================================================================

\section{Regional-scale building portfolio benchmark}
\label{sec:regional_benchmark}
\noindent
This section applies the proposed identification framework to building responses from a smaller earthquake and evaluates how the resulting models affect regional estimates of seismic response, structural damage, and repair cost. An OpenSees MDOF reference portfolio is compared with archetypal and response-informed BW SDOF portfolios for Milpitas, California. All three portfolios share the building inventory, ground-motion inputs, and loss-assessment framework, enabling a comparison of their structural modeling and parameter-assignment approaches.

%-----------------------------------------------------------------------
\subsection{Building inventory}
\label{subsec:regional_inventory}
\begin{figure}[H]
    \centering
    \includegraphics[width=0.8\linewidth, trim=0cm 0.4cm 0cm 0cm, clip]{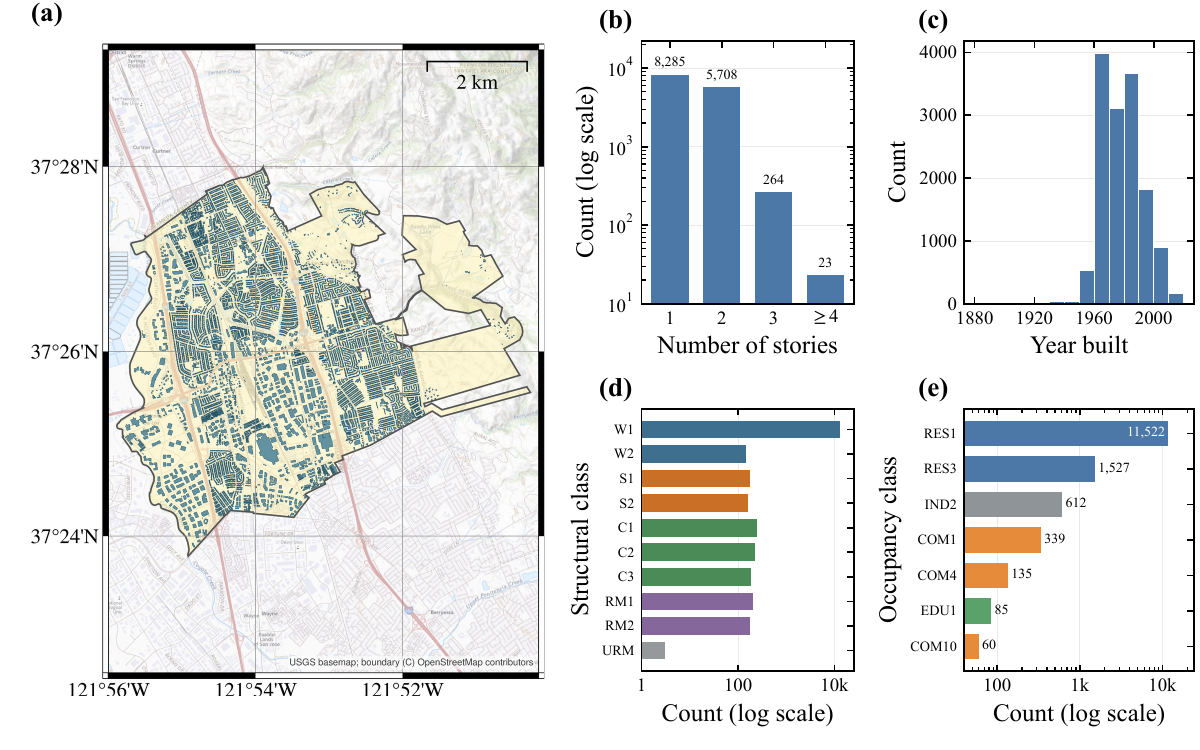}
    \caption{\textbf{Milpitas building inventory.} Panel (a) shows the city boundary and footprints of the 14,280 buildings; panels (b)--(e) summarize story count, construction year, HAZUS structural class, and occupancy, respectively. Colors distinguish structural-material families in panel (d), and occupancy families in panel (e).}
    \label{fig:milpitas_inventory}
\end{figure}

\noindent
The benchmark comprises 14,280 buildings from the Milpitas subset of the NHERI SimCenter Earthquake Testbed for the San Francisco Bay Area~\cite{zsarnoczay_simcenter_2023}. The inventory provides building locations, geometry, construction year, structural type, occupancy, and replacement-cost information. Site $V_{s30}$ values are assigned from the U.S. Geological Survey California $V_{s30}$ dataset~\cite{thompson_updated_2018}. Building heights are calculated by multiplying the number of stories by an assumed story height of $2.5\,\mathrm{m}$. Figure~\ref{fig:milpitas_inventory} summarizes the inventory composition.

%-----------------------------------------------------------------------
\subsection{Ground-motion fields}
\label{subsec:regional_gm}
\noindent
Synthetic acceleration histories are generated using the Southern California Earthquake Center Broadband Platform (BBP)~\cite{maechling_scec_2015} for strike-slip earthquake scenarios on the Hayward fault. The BBP integrates physics-based models of earthquake rupture and seismic-wave propagation with low- and high-frequency synthesis to generate broadband ground motions. The scenario is specified by an epicenter at $37.666^{\circ}\mathrm{N}$, $122.076^{\circ}\mathrm{W}$, strike $325^{\circ}$, dip $90^{\circ}$, rake $180^{\circ}$, and a top-of-rupture depth of $3\,\mathrm{km}$. One $M_w=5.8$ realization is used for model calibration. Evaluation uses 100 BBP realizations at each of four magnitudes, $M_w=5.9$, $6.2$, $6.5$, and $6.8$.

\begin{figure}[H]
    \centering
    \includegraphics[width=0.40\linewidth]{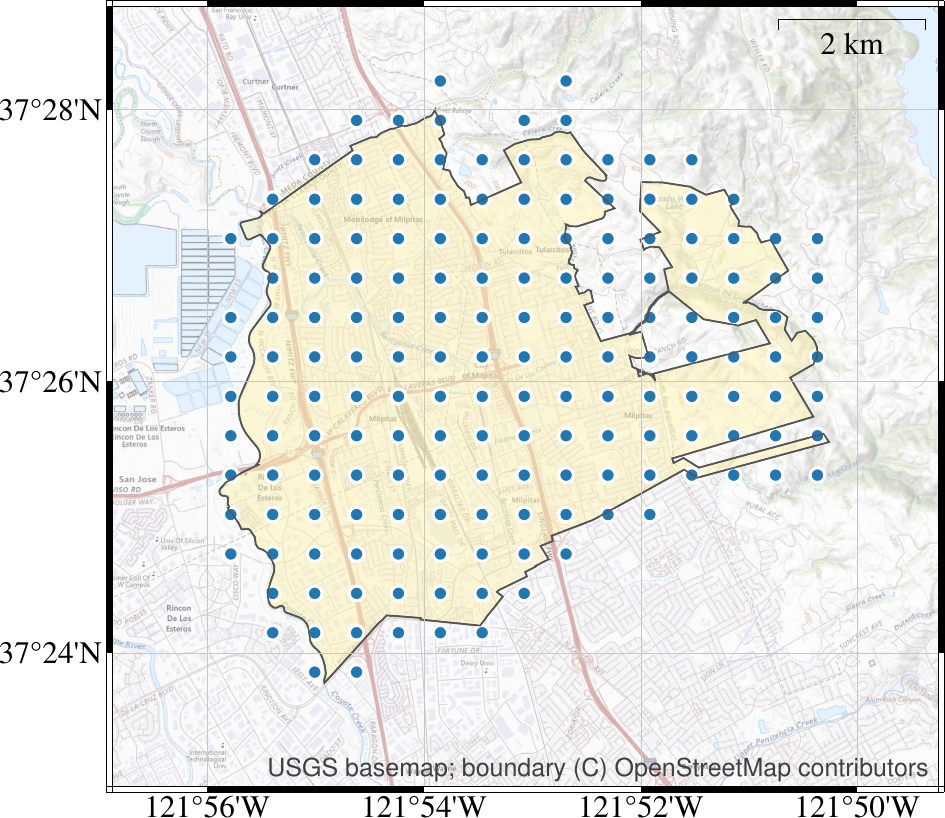}
    \caption{\textbf{BBP station grid for the Milpitas benchmark.} Blue markers show the 168 stations, spaced approximately 550~m apart and extending beyond the city boundary to provide coverage near its limits.}
    \label{fig:milpitas_bbp_stations}
\end{figure}

Each realization provides acceleration histories at 168 grid stations with approximately $550\,\mathrm{m}$ spacing in and around Milpitas (Figure~\ref{fig:milpitas_bbp_stations}). Only the north--south component is used. Each building is assigned to one of its four nearest stations~\cite{mckenna_nheri-simcenterr2dtool_2025}, and this assignment is held fixed across the calibration and evaluation fields and all three portfolios.

%-----------------------------------------------------------------------
\subsection{Structural modeling of the building portfolio}
\label{subsec:regional_portfolio}
\noindent
The building portfolio is represented using three structural modeling approaches, as summarized in Table~\ref{tab:regional_portfolios}. The OpenSees MDOF models, adapted from the framework of~\citet{lu_open-source_2020}, serve as the numerical reference. In this portfolio, story-level parameters are assigned based on the HAZUS capacity curves for the corresponding structural-class/design-level groups~\cite{federal_emergency_management_agency_fema_hazus_2024}. Archetypal BW parameters are obtained by directly fitting the same capacity curves. Response-informed BW parameters, $(T,F_y,\alpha,\beta,n,\zeta)$, are identified separately for each building from its MDOF roof response under the $M_w=5.8$ field.

This calibration setting represents a potential monitoring application in which responses recorded during a smaller earthquake are used to update building models before a subsequent strong event. Such updates are relevant to in-service buildings whose properties may differ from design assumptions due to deterioration or damage. Responses within the elastic range primarily inform stiffness and natural period, while responses extending into yielding may also help identify yield strength. For poorly constrained nonlinear parameters, HAZUS-based values could be retained or used as prior information. In the present example, all six BW parameters were estimated from simulated responses under the $M_w=5.8$ field, without fixing parameters to HAZUS-based values or using those values as priors.

\begin{table}[H]
    \centering
    \small
    \caption{\textbf{Structural modeling and parameter assignment for the regional portfolios.} The archetypal BW and MDOF reference portfolios share the same underlying HAZUS capacity-curve information, which also informs the response-informed BW portfolio indirectly through calibration to the MDOF roof responses. See Appendix~\ref{apdx:portfolio_construction} for details of model construction and parameter assignment.}
    \label{tab:regional_portfolios}\vspace{-0.2cm}
    \begin{tabularx}{0.98\linewidth}{
        @{}
        >{\raggedright\arraybackslash}p{0.21\linewidth}
        >{\raggedright\arraybackslash}p{0.26\linewidth}
        >{\raggedright\arraybackslash}X
        @{}
    }
        \toprule
        Portfolio & Structural modeling & Parameter assignment \\
        \midrule
        MDOF reference
        & OpenSees MDOF shear model
        & HAZUS-based story parameters \\
        \addlinespace
        Archetypal BW
        & BW SDOF
        & Archetype-level HAZUS capacity fit \\
        \addlinespace
        Response-informed BW
        & BW SDOF
        & Building-specific identification from MDOF response \\
        \bottomrule
    \end{tabularx}
\end{table}

\begin{figure}[H]
    \centering
    \includegraphics[width=0.8\linewidth, trim=0cm 0.2cm 0cm 0cm, clip]{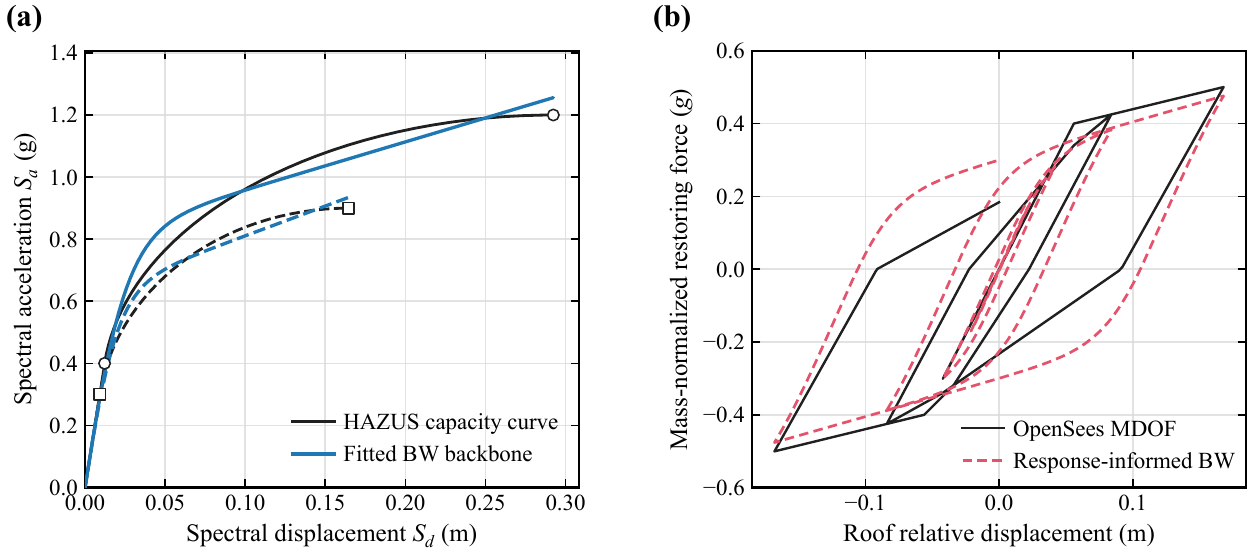}
    \caption{\textbf{Archetypal and response-informed BW model construction.}
    (a) HAZUS capacity curves (black) and fitted archetypal BW backbones (blue) for W1 High-Code (solid) and Moderate-Code (dashed) buildings. Circles and squares mark the yield and ultimate anchors for the High-Code and Moderate-Code curves, respectively.
    (b) Illustrative quasi-static cyclic responses of the OpenSees MDOF reference (black solid) and the calibrated response-informed BW model (red dashed). The BW parameters are identified using the simulated MDOF roof response under the $M_w=5.8$ field.}
    \label{fig:bw_portfolio_construction}
\end{figure}

The archetypal BW models are constructed from the HAZUS capacity curves for each structural-class/design-level group~\cite{fema2024hazus}. Since these curves are expressed in spectral coordinates, the BW responses are computed in the same coordinates. For comparison with the other two portfolios, these responses are converted to equivalent first-mode roof responses using the first-mode participation factor of the corresponding OpenSees MDOF model. The response-informed BW models are identified from the roof responses of the OpenSees MDOF models and represent the response directly in roof coordinates, requiring no modal conversion. Figure~\ref{fig:bw_portfolio_construction} illustrates the archetypal and response-informed BW construction approaches in panels (a) and (b), respectively. All three portfolios inherit HAZUS-based assumptions; the MDOF portfolio serves as a numerical benchmark, and agreement with it does not establish accuracy for the physical buildings. Appendix~\ref{apdx:portfolio_construction} details the portfolio construction and parameter-assignment procedures.

%-----------------------------------------------------------------------
\subsection{Repair-cost assessment}
\label{subsec:regional_cost}
\noindent
Repair costs are estimated within the PELICUN framework using the HAZUS 6.1 damage-and-loss models~\cite{zsarnoczay_pelicun_2020,federal_emergency_management_agency_fema_hazus_2024}. This benchmark considers structural (STR) components. Peak roof drift ratio is adopted as the engineering demand parameter (EDP). HAZUS structural damage-state consequence functions map each sampled damage state to an occupancy-dependent repair-cost ratio. These ratios are multiplied by the building replacement costs provided in the original NHERI SimCenter dataset~\cite{zsarnoczay_simcenter_2023} to obtain monetary losses.

For each building and BBP realization, 1,000 damage-state samples are generated conditional on the computed EDP and converted to structural repair-cost samples. Building-level samples are summed across the portfolio to obtain regional cost samples. Combining the 1,000 regional samples from each of the 100 BBP realizations yields $100\times1{,}000=100{,}000$ samples per magnitude and portfolio. These samples represent variability across the selected ground-motion fields and conditional damage-state uncertainty at the computed EDPs. All three portfolios use the same fragility and consequence models and sampling settings.

%-----------------------------------------------------------------------
\subsection{Regional damage and repair cost assessment}
\label{subsec:regional_results}
\begin{figure}[H]
    \centering
    \includegraphics[width=0.9\linewidth]{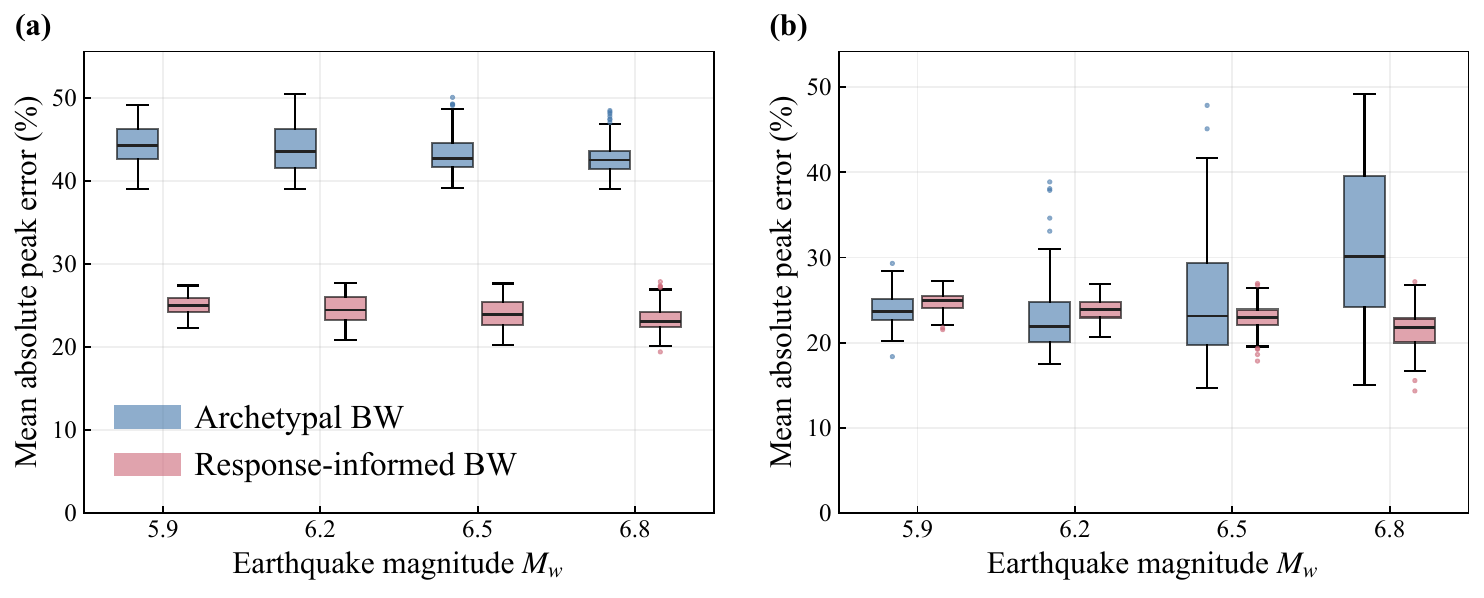}
    \vspace{-0.3cm}
    \caption{\textbf{Portfolio-mean errors in peak structural response relative to the MDOF reference.} Panels (a) and (b) show errors in peak roof drift ratio and peak absolute roof acceleration, respectively. Blue boxes denote the archetypal BW models and red boxes denote the response-informed BW models. Each box summarizes 100 BBP realizations at the indicated magnitude. For each realization, building-level absolute percentage errors are averaged over the common portfolio used in the damage and repair-cost comparisons. Boxes span the interquartile range, center lines denote medians, whiskers extend to the most extreme values within 1.5 interquartile ranges, and points show the remaining observations.}
    \label{fig:regional_portfolio_mean_error_boxplot}
\end{figure}

\noindent
Figure~\ref{fig:regional_portfolio_mean_error_boxplot} first compares the BW portfolios' peak response errors relative to the MDOF reference under the larger earthquake scenarios following calibration to the $M_w=5.8$ field. The response-informed models consistently reduce median portfolio-mean drift errors from $42.5$--$44.3\%$ to $23.1$--$25.0\%$ across the four magnitudes. Improvements in acceleration accuracy are less consistent across magnitudes, while the reduction in drift errors is directly relevant to the subsequent structural damage and repair-cost assessment.

\begin{figure}[H]
    \centering
    \includegraphics[width=0.9\linewidth]{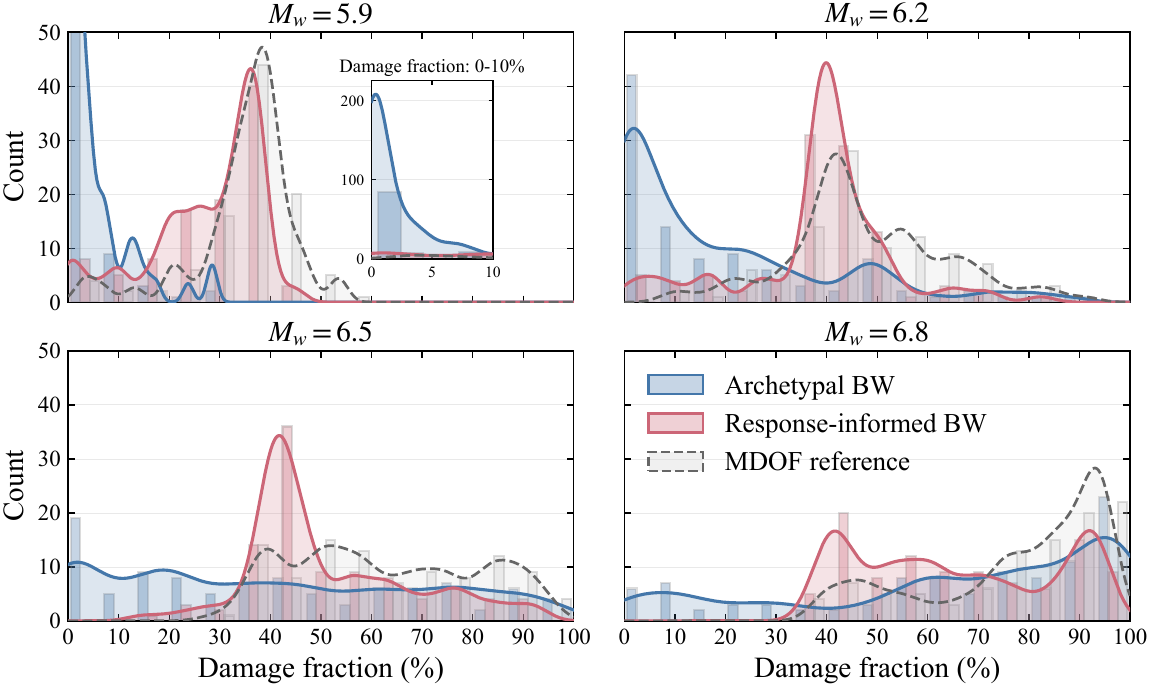}
    \caption{\textbf{Regional distributions of DS1 median-threshold exceedance.} Histograms summarize 100 BBP realizations at each magnitude for the archetypal BW (blue), response-informed BW (red), and MDOF reference (gray) portfolios. Curves are kernel density estimates scaled to the histogram count axis; the reference curve is dashed. The damage fraction is the percentage of the common 14,280 buildings whose peak roof drift ratio reaches or exceeds the building-specific median structural DS1 threshold. The inset for $M_w=5.9$ displays the full vertical range within the $0$--$10\%$ damage-fraction interval.}
    \label{fig:regional_damage_combined}
\end{figure}

Figure~\ref{fig:regional_damage_combined} shows that building-specific updating substantially changes the damage-fraction distributions. For each BBP realization, the structural damage fraction is defined as the percentage of buildings whose peak roof drift ratios reach or exceed their assigned median structural DS1 drift thresholds from HAZUS~6.1~\cite{fema2024hazus}. At $M_w=5.9$, the archetypal distribution is concentrated near zero, whereas the response-informed distribution captures the dominant reference concentration around $35$--$40\%$. Agreement is less complete at larger magnitudes. At $M_w=6.5$, the response-informed distribution is concentrated around $40$--$50\%$, while the reference spans a wider range with more realizations at higher damage fractions. At $M_w=6.8$, the archetypal and response-informed mean damage fractions are nearly identical ($64.9\%$ and $65.5\%$), but both are below the reference mean of $77.4\%$ and their distribution shapes differ. Although the response-informed mean is closer to the reference at every magnitude, the improvement does not extend uniformly to all features of the distributions. These differences show that similar average damage levels can conceal distinct patterns of concentration and dispersion across ground-motion realizations. The sensitivity of these patterns to structural modeling is consistent with recent findings that coarse structural categorization can distort the description of collective risk~\cite{oh2026phase}.

\begin{figure}[H]
    \centering
    \includegraphics[width=0.9\linewidth, trim=0cm 2.3cm 0cm 0.3cm, clip]{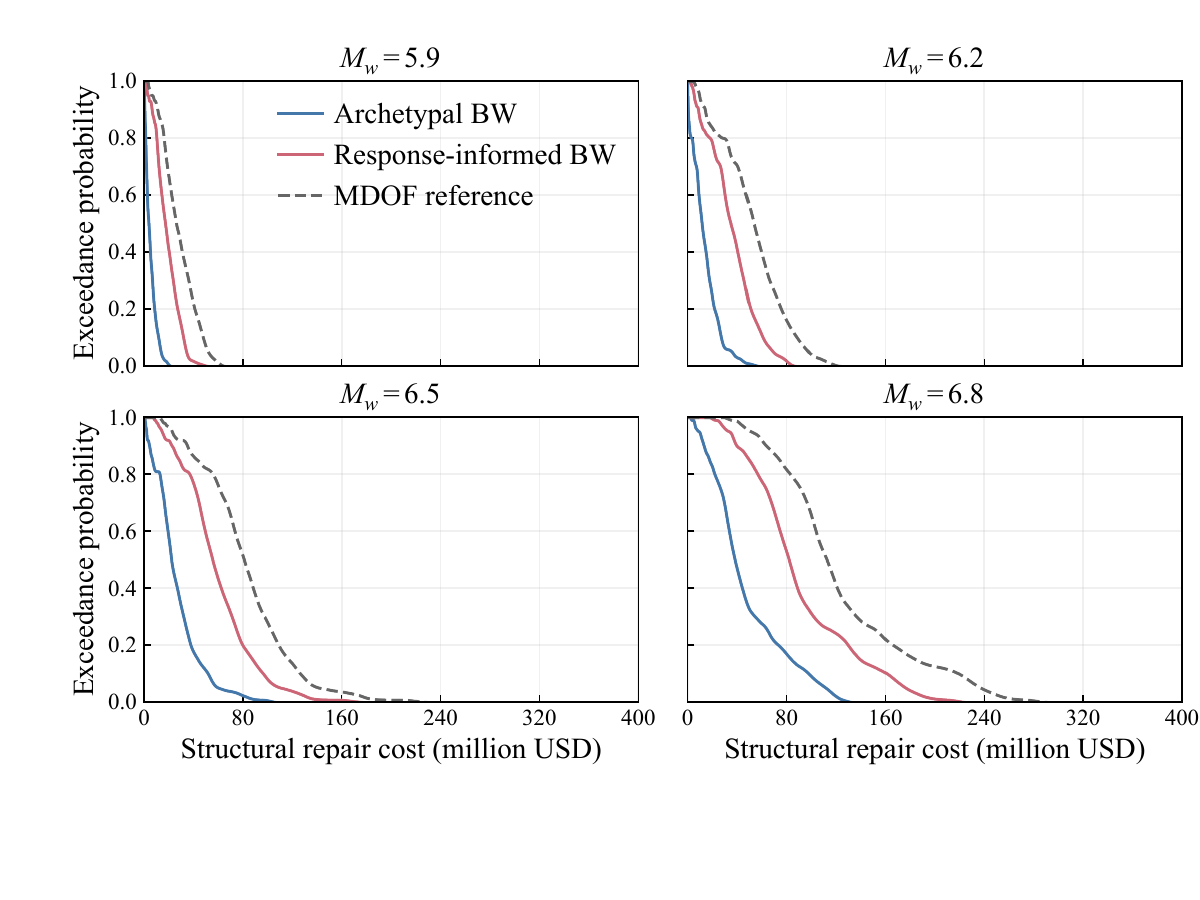}
    \caption{\textbf{Exceedance curves of regional structural repair cost.} Blue and red solid curves denote the archetypal and response-informed BW portfolios, respectively; gray dashed curves denote the MDOF reference. Each magnitude uses the same 14,280 buildings and 100,000 regional cost samples per modeling approach, obtained from 100 BBP realizations with 1,000 conditional damage-state samples each. Costs include only the structural (STR) repair component. Exceedance probabilities are conditional on the selected scenario ensemble and are not annual probabilities.}
    \label{fig:regional_repair_cost_combined}
\end{figure}

Figure~\ref{fig:regional_repair_cost_combined} compares the exceedance curves of regional structural repair cost for all three portfolios. Each curve gives the fraction of the 100,000 regional cost samples exceeding a given cost level. These probabilities are conditional on the selected scenario ensemble and do not represent annual exceedance probabilities. The response-informed curves shift toward the MDOF reference relative to the archetypal curves at all four magnitudes, indicating higher anticipated structural repair costs after building-specific updating. Mean structural repair costs, obtained by averaging the regional samples, are closer to the reference for the response-informed models at every magnitude. At $M_w=6.8$, the archetypal, response-informed, and reference means are 47.4, 91.7, and 125.4 million USD, respectively. Across the four magnitudes, the archetypal means are approximately $62$--$83\%$ below the reference, compared with $27$--$35\%$ for the response-informed models. Building-specific updating therefore reduces the discrepancy in structural repair costs, although substantial underestimation remains. These results indicate that building-specific updating brings regional repair-cost estimates closer to the MDOF reference, although appreciable underestimation remains across all four magnitudes. The improvement is therefore relative to the archetypal baseline and does not imply full agreement with the reference cost distributions.

The regional comparison demonstrates that building-specific response information can materially change the anticipated extent of structural damage and the resources required for structural repair. Although discrepancies from the MDOF reference remain, partly attributable to the equivalent SDOF representation of multistory behavior, incorporating this information substantially reduces errors in mean structural repair costs relative to the archetypal models. Even when average damage fractions are similar, the predicted damage distributions and structural repair costs can differ substantially. These results emphasize the importance of incorporating building-specific response information into regional structural damage and repair-cost assessment, even within simplified structural models.

%=======================================================================
% Conclusions
%=======================================================================
\section{Conclusions}
\label{sec:conclusions}
\noindent
This study developed a modular physics-regularized framework for identifying building-specific equivalent single-degree-of-freedom (SDOF) Bouc--Wen models from seismic input--response histories. A differentiable forward solver enables joint estimation of hysteretic parameters and viscous damping through an objective combining response matching with independently weighted dynamic-equilibrium and hysteretic state-evolution residuals. Automatic differentiation through the time-integration scheme provides gradients for parameter estimation, while an integral formulation of the state-evolution residual avoids numerical differentiation of the reconstructed hysteretic state. The modular objective accommodates different response-matching channels and allows the contributions of the physics residuals to be adjusted independently.

For 100 synthetic Bouc--Wen model with degradation systems, the gradient-based algorithm achieved lower median parameter and response prediction errors than a genetic algorithm using the same observations and objective under comparable elapsed fitting times. The hybrid-simulation example extended the evaluation to experimentally informed multistory responses, demonstrating improved calibration and prediction under an unseen earthquake record with physics regularization. A complementary loss ablation study showed reduced variability across calibration records and suppression of large prediction errors, although median prediction error remained nearly unchanged. These results support the use of gradient-based identification and physics regularization to construct building-specific models for response prediction beyond the calibration record. Then, Milpitas numerical benchmark showed that updating structural models from simulated responses to a smaller earthquake can significantly change anticipated regional damage and structural repair needs under larger scenarios. With the building inventory, ground-motion inputs, and damage-and-loss models held fixed, the archetypal and response-informed portfolios yielded distinct damage-fraction distributions and substantially different mean structural repair costs. These results highlight the value of building-specific response information for regional damage and repair-cost assessment even within simplified structural models. These findings motivate integrated frameworks that connect advances in structural response sensing and measurement with building-specific nonlinear hysteretic modeling and regional risk assessment. 

Further research is needed to establish the framework's applicability to operational regional assessments. This includes extending the equivalent SDOF formulation to multimodal or multi-degree-of-freedom representations and distinguish the effects of structural idealization from parameter estimation errors on regional loss predictions. Broader experimental and field validation should address realistic sensor configurations, measurement noise, missing response channels, and uncertainty in the quantities reconstructed for the physics residuals. Particular attention should be given to parameter identifiability under limited nonlinear excitation, since a good calibration fit alone does not establish reliable prediction under stronger shaking. Evaluation across a broader range of structures, calibration records, and shaking intensities would help establish these predictive limits. Future work should also develop systematic procedures for selecting regularization weights and assess sensitivity to parameter initialization within practical computational budgets. Finally, propagating uncertainties in the identified parameters and model form through damage-and-loss calculations and assessing sensitivity to HAZUS-based assumptions would help quantify uncertainty in regional repair-cost estimates and evaluate the robustness of decisions based on them.

%=======================================================================
% Acknowledgments
%=======================================================================
\section*{Acknowledgments}\noindent
This study was supported by Korea Institute for Advancement of Technology (KIAT) funded by the Ministry of Trade, Industry and Energy in 2025 (Grant Number: P0030038). We thank Jinyan Zhao for generating the SCEC Broadband Platform (BBP) ground-motion simulations used in this study. This research used the Savio computational cluster resource provided by the Berkeley Research Computing program at the University of California, Berkeley (supported by the UC Berkeley Chancellor, Vice Chancellor for Research, and Chief Information Officer).

%=======================================================================
% References
%=======================================================================
\bibliography{references}

%=======================================================================
% Appendix
%=======================================================================
\newpage
\appendix

\counterwithin*{figure}{section}
\counterwithin*{table}{section}

\renewcommand{\thefigure}{\Alph{section}.\arabic{figure}}
\renewcommand{\thetable}{\Alph{section}.\arabic{table}}

%--------------------------------------------------
\section{Loss ablation study}
\begin{figure}[H]
    \centering
    \includegraphics[width=0.4\linewidth, trim=0cm 0.2cm 0cm 0cm, clip]{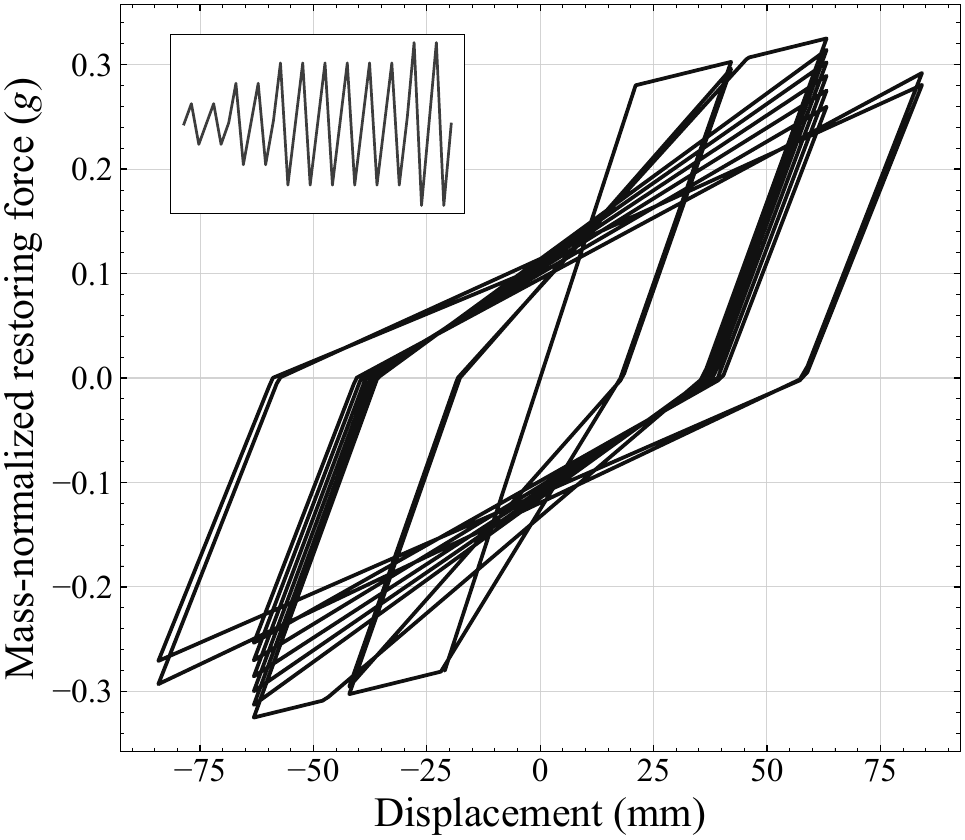}
    \caption{\textbf{Cyclic response of the OpenSees reference model.} Restoring force is normalized by mass and expressed in $g$. The inset shows the prescribed displacement history for this quasi-static illustration.}
    \label{fig:opensees_load_disp_loop}
\end{figure}

\appendixletterlabel{apdx:loss_ablation}
\noindent
This appendix examines how the dynamic-equilibrium and state-evolution losses affect parameter estimates, prediction accuracy, and variability across calibration records under model-form mismatch. The OpenSees reference is a unit-mass SDOF oscillator using a \texttt{zeroLength} element with the \texttt{Hysteretic} material, a symmetric bilinear envelope, cyclic deterioration, and unloading-stiffness degradation. Its hysteretic formulation differs from that of BWdeg, allowing assessment under model-form mismatch. Table~\ref{tab:opensees_reference_model} lists the model parameters, and Figure~\ref{fig:opensees_load_disp_loop} shows its cyclic response.

\begin{table}[H]
    \centering
    \caption{\textbf{Parameters of the OpenSees reference SDOF model.}}
    \label{tab:opensees_reference_model}\vspace{-0.2cm}
    \setlength{\tabcolsep}{6pt}
    \begin{tabular}{@{}lcc@{}}
        \hline
        Parameter & Value & Unit \\
        \hline
        Mass                              & $1.0$   & kg \\
        Initial period                    & $0.55$  & s  \\
        Damping ratio                     & $0.05$  & -- \\
        Yield acceleration                & $0.28$  & $g$ \\
        Post-yield stiffness ratio        & $0.08$  & -- \\
        Ductility-based damage parameter  & $0.04$  & -- \\
        Energy-based damage parameter     & $0.005$ & --\\
        Unloading-stiffness degradation exponent & $0.10$ & -- \\
        \hline
    \end{tabular}
\end{table}

A total of 500 ground-motion records are drawn from NGA-West2~\cite{ancheta_nga-west2_2014}, of which 50 are selected for calibration and the remaining 450 are reserved for prediction tests. The calibration records are selected based on the OpenSees reference responses, without considering fitting errors, to provide repeated inelastic cycles and observable deterioration while limiting amplitude scaling. They are scaled to a target displacement ductility of $\mu=3.0\pm0.05$ using scale factors no greater than $2$, where $\mu=\max_t|u(t)|/u_y$ and $u_y$ is the reference yield displacement. These records produce at least four inelastic reversals, sufficient hysteretic energy dissipation, and measurable cyclic degradation.

Each calibration record is used in a separate identification run for each loss configuration, using displacement, velocity, and relative acceleration histories as observations. All fits start from the same initial values: $T=1.833$~s, $F_y=0.879g$, $\alpha=0.167$, $\beta=0.279$, $n=2.220$, $\delta_\nu=0.0984~\mathrm{s^2/m^2}$, $\delta_\eta=0.359~\mathrm{s^2/m^2}$, and $\zeta=0.0585$. Each identified model is then evaluated against the OpenSees responses to the 450 unseen records at their original amplitudes. All loss configurations use the same calibration and evaluation sets.

\begin{figure}[H]
    \centering
    \includegraphics[width=0.9\linewidth, trim=0cm 0.3cm 0cm 0cm, clip]{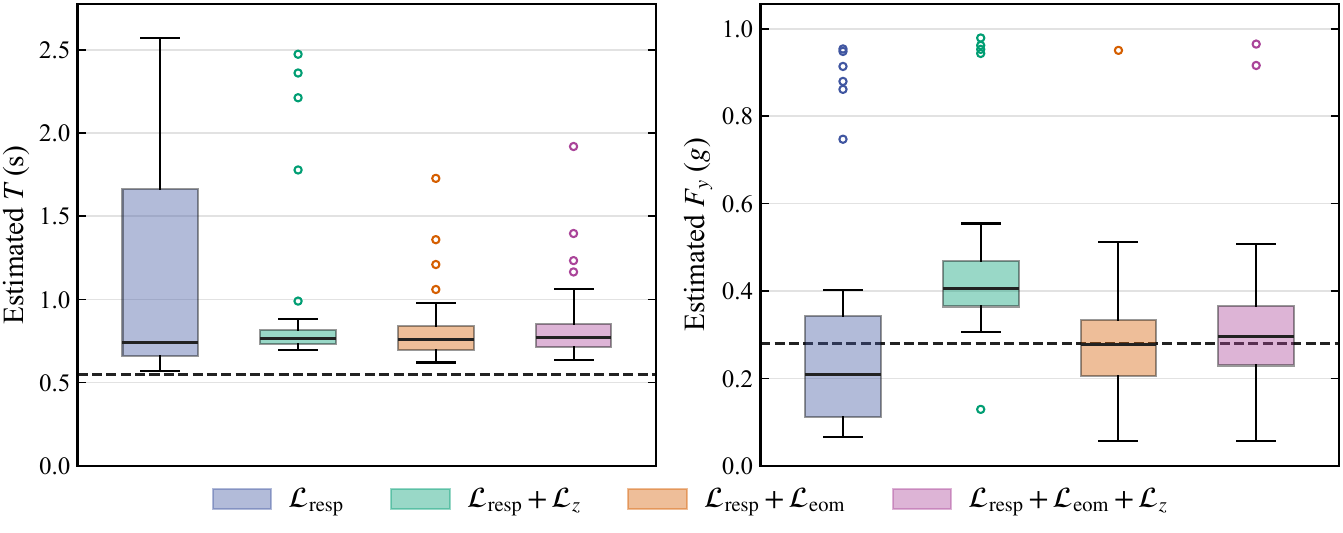}
    \caption{\textbf{Comparison of key parameter estimates with the OpenSees reference: period and yield strength.} Left and right panels show $T$ and $F_y$ after 2,000 epochs for the four loss configurations. Each box summarizes estimates from separate calibration records. Dashed lines indicate the reference values, $T=0.55$~s and $F_y=0.28g$, with $F_y$ expressed as yield acceleration for the unit-mass system. Boxes span the interquartile range (IQR), center lines denote medians, whiskers extend to the most extreme values within 1.5 IQR of the box, and circles denote outliers.}
    \label{fig:opensees_reference_parameter_comparison}
\end{figure}

Including the equilibrium residual substantially reduces the variability of estimated periods and yield strengths across calibration records (Figure~\ref{fig:opensees_reference_parameter_comparison}). The median $F_y$ estimates are $0.277g$ for $\mathcal{L}_{\mathrm{resp}}+\mathcal{L}_{\mathrm{eom}}$ and $0.296g$ for the full objective, close to the reference value of $0.28g$. Period estimates are more concentrated, although the full-objective median of approximately $0.77$~s remains above the reference value of $0.55$~s. Although $T$ is defined by the initial elastic stiffness in both the OpenSees and BWdeg models, differences in their yielding, unloading, and degradation laws can lead to compensating adjustments in stiffness, damping, and hysteretic parameters when fitting nonlinear response histories. These adjustments may partly explain the period offset, although incomplete convergence within the prescribed number of epochs may also contribute.

\begin{figure}[H]
    \centering
    \includegraphics[width=0.9\linewidth, trim=0cm 0.3cm 0cm 0cm, clip]{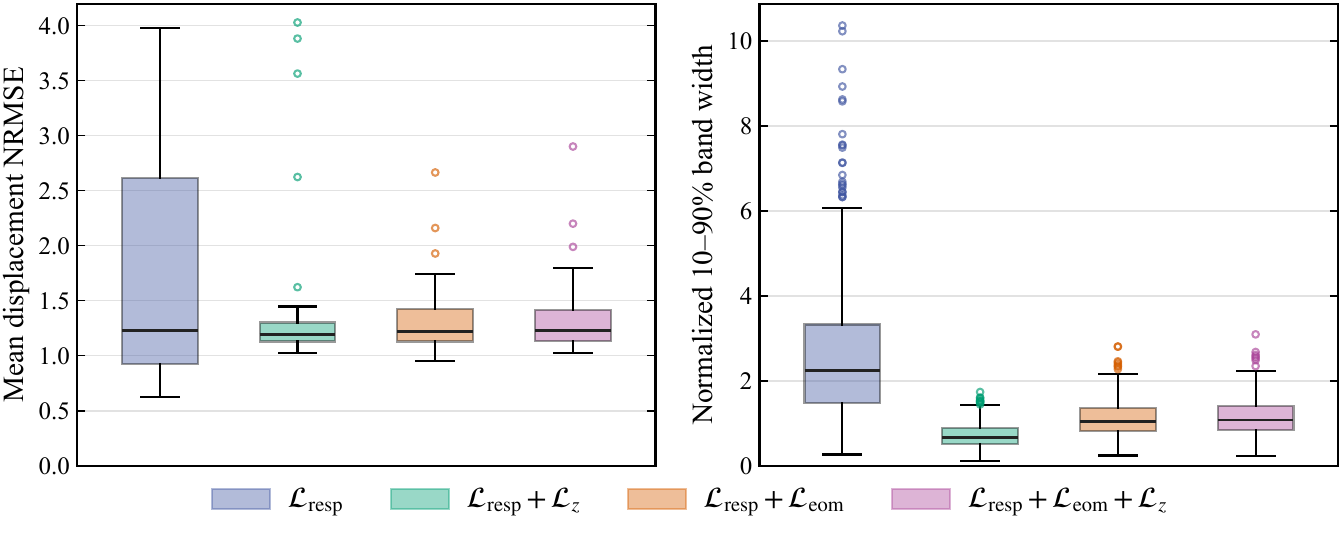}
    \caption{\textbf{Unseen displacement prediction accuracy and variability across calibration records.} The left panel shows the mean displacement NRMSE for each identified model, averaged over 450 unseen records. The right panel shows the time-averaged pointwise 10--90\% displacement band width across identified models for each unseen record. RMSE and band width are normalized by the corresponding reference displacement RMS and shown as dimensionless ratios. Lower values indicate greater prediction accuracy in the left panel and less variability in the right panel. Boxplot conventions follow Figure~\ref{fig:opensees_reference_parameter_comparison}.}
    \label{fig:displacement_accuracy_spread}
\end{figure}

Physics regularization primarily reduces the spread of displacement prediction errors across calibration fits, as shown in Figure~\ref{fig:displacement_accuracy_spread} for the unseen ground motions. Relative to response-only calibration, the full objective reduces the ensemble mean of the per-fit mean NRMSE by approximately 19\% and the median normalized prediction-band width by approximately 52\%, while leaving the median NRMSE nearly unchanged. Adding only the state-evolution loss gives the lowest median error and narrowest prediction bands, but retains several fits with large mean errors. Objectives containing the equilibrium residual more effectively limit these large errors, with the full objective performing similarly to response matching with dynamic equilibrium.

\begin{figure}[H]
    \centering
    \includegraphics[width=0.8\linewidth, trim=0cm 0.1cm 0cm 0cm, clip]{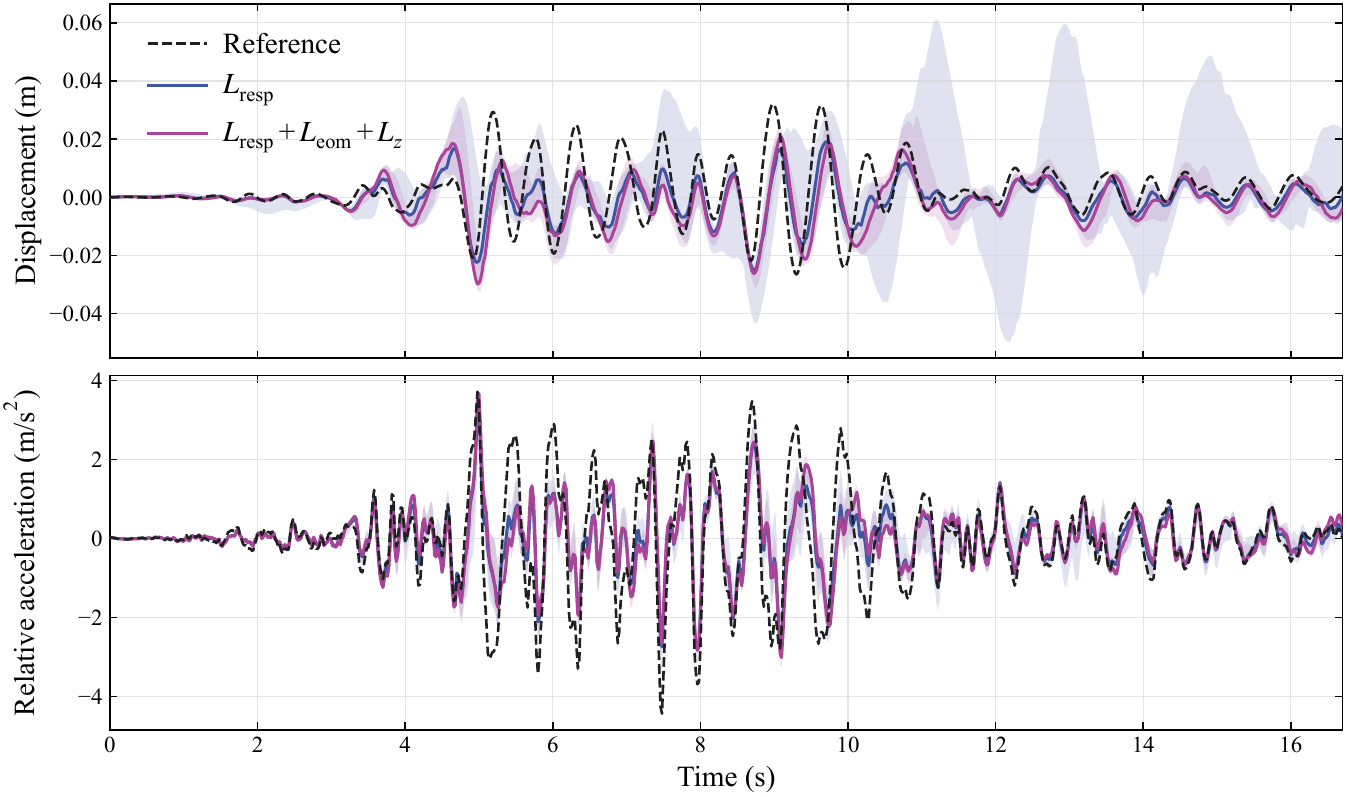}
    \caption{\textbf{Effect of physics regularization on response variability under an unseen ground motion.} Upper and lower panels show relative displacement and acceleration for response-only calibration and the full objective after 2,000 epochs. Dashed curves denote the OpenSees reference; solid curves and shaded bands show the pointwise median and 10--90\% range of responses predicted using the 50 parameter sets identified from separate calibration records for each loss configuration. Bands represent variability across calibration records, not confidence intervals.}
    \label{fig:unseen_response_time_history}
\end{figure}

Figure~\ref{fig:unseen_response_time_history} illustrates the response to one unseen ground motion. Compared with response-only calibration, the full objective narrows the displacement prediction band, particularly in the later part of the record. Both configurations capture the main response oscillations, but displacement amplitude and phase discrepancies and underestimated acceleration peaks remain. Together, the results show that physics regularization improves consistency across calibration records under model-form mismatch, while more concentrated predictions do not necessarily imply closer agreement with the reference history.

%--------------------------------------------------
\section{Structural portfolio construction}
\appendixletterlabel{apdx:portfolio_construction}
\noindent
The three portfolios summarized in Table~\ref{tab:regional_portfolios} use the same building inventory, ground-motion fields, and building-to-station assignments described in Section~\ref{sec:regional_benchmark}. This appendix documents the portfolio-specific structural representations, parameter-assignment procedures, and analysis settings.

\subsection{OpenSees MDOF reference portfolio}
\appendixletterlabel{apdx:mdof_reference_portfolio}
\noindent
Each building is represented by an OpenSees shear-building model with lumped floor masses. Two uncoupled, identical translational degrees of freedom are defined at each floor, floor rotation is restrained, and uniform base excitation is applied in the north--south direction used in the benchmark. Story height, floor mass, and the positive backbone parameters are assigned from the HAZUS-based mechanical inventory according to structural class and design era; the negative backbone branch is assigned symmetrically. Each story is implemented as a zero-length element with the OpenSees \texttt{Hysteretic} uniaxial material~\cite{McKenna2011OpenSees:Simulation,zhu_openseespy_2018}. The inventory parameters \texttt{pinch} and \texttt{beta} control pinched reloading and unloading-stiffness deterioration, respectively, while the explicit ductility- and energy-damage coefficients are set to zero.

A building-specific damping ratio is sampled uniformly between 0.02 and 0.05 using the building identifier to ensure reproducibility and is applied to all translational modes of that building. The assigned BBP record is imposed through \texttt{UniformExcitation}. Transient response is evaluated with the average-acceleration Newmark method, $(\gamma,\beta_N)=(0.5,0.25)$, using a displacement-increment convergence tolerance of $10^{-8}$ and a maximum of 30 iterations. Newton, Newton line-search, and modified-Newton algorithms are attempted sequentially when required for convergence. The retained outputs are roof relative displacement, velocity, and acceleration; roof absolute acceleration; peak roof drift ratio; peak interstory drift ratio; and peak floor absolute acceleration. The MDOF portfolio contains 14,280 buildings and is analyzed under the calibration and evaluation ground-motion fields described in Section~\ref{sec:regional_benchmark}.

\subsection{HAZUS-based archetypal BW portfolio}
\appendixletterlabel{apdx:archetypal_bw_portfolio}
\noindent
For the class-based baseline, the HAZUS capacity curve associated with each structural archetype and design level is converted from spectral displacement--spectral acceleration coordinates to a unit-mass BW model. If $(D_y,A_y)$ denote the HAZUS yield point in consistent units, the initial stiffness and period are
\begin{equation}
    k_0 = \frac{A_y}{D_y},
    \qquad
    T = 2\pi\sqrt{\frac{D_y}{A_y}}.
    \label{eq:hazus_initial_period}
\end{equation}
With $T$ fixed, $F_y$, $\alpha$, and $n$ are obtained by bounded least squares against 240 points spanning the complete monotonic HAZUS curve from the origin to the ultimate displacement, with additional residuals imposed at the yield and ultimate anchors. Because the monotonic capacity curve does not identify cyclic shape or viscous damping, these parameters are fixed at $\beta=0.5$ and $\zeta=0.05$, respectively; degradation parameters are omitted. This procedure produces 32 distinct BW parameter sets for the inventory. Buildings in the same HAZUS archetype/design-level group receive the same parameter vector, so within-group differences arise from location and ground-motion input rather than from the structural model.

The archetypal BW equations are integrated in the HAZUS spectral coordinate. Before response comparison and EDP calculation, displacement, velocity, and relative acceleration are reconstructed in an equivalent first-mode roof coordinate as $u_r(t)=\Gamma_1\phi_{r1}u(t)$ and analogously for the time derivatives. The elastic mode shape and participation factor are calculated from the initial mass and stiffness matrices of the corresponding mechanical MDOF model and normalized such that $\phi_{r1}=1$; ground acceleration is added after scaling the relative acceleration. This coordinate mapping uses no earthquake-response history. The response-informed SDOF, by comparison, is identified directly in the MDOF roof-response coordinate and therefore receives no modal rescaling.

\subsection{Response-informed BW portfolio}
\appendixletterlabel{apdx:response_informed_portfolio}
\noindent
The response-informed portfolio is constructed by applying the single-record procedure of Section~\ref{sec:identification} independently to the OpenSees roof response of every building under the $M_w=5.8$ identification field. Roof relative acceleration is used as the response-matching channel, and the complete physics-regularized objective estimates $(T,F_y,\alpha,\beta,n,\zeta)$ for a unit-mass BW model. Ground acceleration and the saved roof displacement, velocity, and acceleration enter the physics losses as described in Section~\ref{subsec:objective}; each record is cropped according to Section~\ref{subsec:implementation}. The same parameter bounds, weak-form window, optimizer settings, and 200-epoch budget are used for all buildings.

The equivalent-SDOF coordinate is defined directly by the MDOF roof relative response,
\begin{equation}
    u_i^{\mathrm{obs}}(t)=u_{r,i}^{\mathrm{MDOF}}(t),
    \qquad
    \dot{u}_i^{\mathrm{obs}}(t)=\dot{u}_{r,i}^{\mathrm{MDOF}}(t),
    \qquad
    \ddot{u}_i^{\mathrm{obs}}(t)=\ddot{u}_{r,i}^{\mathrm{MDOF}}(t).
    \label{eq:roof_response_coordinate}
\end{equation}
No analytical mode shape, participation factor, or effective modal mass is imposed. The unit-mass BW model is therefore treated as an effective roof-coordinate input--output surrogate whose predictive performance is evaluated against the MDOF response (see Figure~\ref{fig:regional_portfolio_mean_error_boxplot}), rather than as an independently verified first-mode physical representation. Both SDOF portfolios are subsequently analyzed under 100 BBP realizations at each of four magnitudes, $M_w=5.9,6.2,6.5,$ and $6.8$, using the same vectorized fixed-step Runge--Kutta solver and the same building-to-station assignments as the MDOF reference.

% \end{linenumbers}
\end{document}